\documentclass{aa}  
\usepackage[hidelinks]{hyperref}
\usepackage{subcaption}
\usepackage{textgreek}
\hypersetup{
  colorlinks   = true, 
  urlcolor     = blue, 
  linkcolor    = blue, 
  citecolor   = blue 
}

\usepackage{graphicx}
\usepackage{listings}
\usepackage{float}
\usepackage{txfonts}

\begin{document}

   \title{Star formation and accretion rates within 500 pc as traced by \textit{Gaia} DR3 XP spectra}

    \author{L. Delfini
           \inst{1,2,3}\thanks{\email{lavinia.delfini@outlook.com}}
           \and
           M. Vioque
           \inst{1}
           \and Á. Ribas
           \inst{2}
           \and S. Hodgkin
           \inst{2}}
    \institute{European Southern Observatory, Karl-Schwarzschild-Str. 2, 85748 Garching bei München, Germany
    \and
    Institute of Astronomy, University of Cambridge, Madingley Road, Cambridge CB3 0HA, United Kingdom
    \and
    European Space Agency (ESA), European Space Astronomy Centre (ESAC), Camino Bajo del Castillo s/n, 28692 Villanueva de la Ca\~nada, Madrid, Spain}

   \date{Received 20 December 2024 / Accepted 2 May 2025}
 
  \abstract
   {Accretion rates from protoplanetary discs onto forming stars are a key ingredient in star formation and protoplanetary disc evolution. Extensive efforts surveying different individual star-forming regions with spectroscopy and narrow-band photometry have been made to derive accretion rates on large populations of young stellar objects (YSOs). }
   {We use \textit{Gaia} DR3 XP spectra to perform the first all-sky homogeneous analysis of YSO accretion properties within 500\,pc.}
   {We characterise the H\textalpha{} line emission of YSOs within 500\,pc by using the H\textalpha{} pseudo-equivalent widths and XP spectra provided by \textit{Gaia} DR3. We derive accretion luminosities and mass accretion rates, together with stellar parameters, for $145\,975$ all-sky candidate YSO H\textalpha{} emitters. We describe filtering strategies to select specific sub-samples of YSOs from this catalogue.}
   {We identify a large population of low-accreting YSO candidates untraced by previous accretion rates surveys. We find previous surveys have mostly focused on YSO populations with significant infrared excess from disc emission. The population of low-accreting YSOs is mostly spatially dispersed, away from star-forming regions or the more clustered environments of star formation. Many YSOs appear entirely disconnected from young populations, and they are reminiscent of the long-lived ‘Peter Pan’ YSOs. We find $L_{\text{acc}}\propto L_\star ^{1.41 \pm 0.02}$ and $\dot M_{\text{acc}} \propto M_{\star}^{2.4 \pm 0.1}$ for the purest all-sky sample of YSO candidates. By fitting an exponential function to the fraction of accreting stars in clusters of different ages in the Sco-Cen complex, we obtain an accretion timescale of $\tau_{\text{acc}} = 2.7 \pm 0.4$\,Myr. The percentage of accretors found by fitting a power law function is $70\%$ at 2\,Myr and 2.8\% at $10$\,Myr.}
   {With this new catalogue of H\textalpha{} emitters, we significantly increase the number of YSO candidates with accretion rate estimations in the local neighbourhood. This allows us to study accretion timescales and the spatial and physical properties of YSO accretion from a large, all-sky, and homogeneous sample for the first time.}


   \maketitle

\section{Introduction}

In forming stars, or young stellar objects (YSOs), the accretion of material from the protoplanetary disc onto the star is regulated by the magnetospheric accretion mechanism (see \citealp{2016ARA&A..54..135H} and references therein). The variability of the accretion process and its characteristic timescales are still objects of intense study (\citealp{2023ASPC..534..355F}), but it is clear that accretion is highly variable (e.g. \citealp{2022A&A...664L...7C,2024ApJ...970..118W}), often happening in small irregular bursts (e.g. \citealp{2018AJ....156...71C}) and on occasion in large outbursts (e.g. \citealp{2021ApJS..256...30K}). A direct measurement of accretion rates can be obtained from the continuum UV excess and the veiling of photospheric emission caused by the accretion shocks onto the stellar photosphere. This UV excess can be observed over the Balmer jump and it has been modelled to derive accretion rates in many nearby YSOs  (e.g. \citealp{2020A&A...639A..58M}). Such a derivation of accretion rates from UV excess emission has also been possible for intermediate-mass YSOs (also called `Herbig' stars, \citealp{Herbigs_bonafide}), although magnetospheric accretion is likely not the dominant accretion mechanism for the more massive YSOs ($M>4$ M$_{\odot}$, \citealp{2020Galax...8...39M, Vioque_2022}).

Empirical correlations have been found between the directly measured accretion traced from continuum UV excess and emission lines in the optical and near-infrared part of the spectrum (e.g. \citealp{Alcala_lupus2,2017MNRAS.464.4721F}). The physical interpretation of these relations is not trivial, as the lines arise from different regions within the YSO environment and often have multiple components (\citealp{2015MNRAS.452.2837M,2018A&A...609A..87N}). Nevertheless, line emission has permitted the derivation of accretion rates in objects too embedded to be observed at UV wavelengths (showing that magnetospheric accretion also dominates at early stages of star formation, or the class I phase, \citealp{2023ApJ...944..135F,2024ApJ...972..149F}), or in samples without UV coverage (e.g. \citealp{Wichittanakom_2020,2023ApJ...945..112F,2024A&A...684L...8R}). In particular, the H\textalpha{} (656.28 nm) Balmer emission line has been extensively used to derive accretion rates as it correlates decently well with the accretion rates measured from the continuum UV excess (\citealp{Alcala_lupus2}), with a scatter of a factor of a few. While other lines are less influenced by non-accreting phenomena and show better correlations with directly measured accretion (e.g. He I, Br\textgamma{}), the H\textalpha{} line has the advantage that it is often the strongest line in the optical spectra of YSOs, and as such it is often the easiest to trace. In fact, H\textalpha{} emission alone has been successfully used to identify populations of YSOs lacking complementary observations (e.g. \citealp{2020A&A...638A..21V}).

In addition to characterising the transfer of material from discs to forming stars, accretion rates are a powerful tool to trace protoplanetary disc evolution (\citealp{Fedele10_accretion_timescale}, \citealp{2016A&A...591L...3M,2023AJ....166..147G}), disc dispersal mechanisms (\citealp{2022MNRAS.512L..74T,2024arXiv241107227Z}), and the impact of the environment (\citealp{2022EPJP..137.1132W,2024A&A...691A.169W}) and of binarity (\citealp{2022MNRAS.512.3538Z}) in star formation. Extensive efforts and several surveys have been dedicated to measuring accretion rates in populations of YSOs across different star-forming regions (see \citealp{Manara_TTauri} and references therein), often targeting entire individual star-forming regions with single-slit spectrographs or narrow-band H\textalpha{} photometry (see \citealp{2010ApJ...715....1D}). 

In this work, we make use of the \textit{Gaia} XP spectra to trace H\textalpha{} emission homogeneously within 500\,pc, using it to describe the accretion properties of the local volume of YSOs. \textit{Gaia} is an European Space Agency's mission providing astrometry, and optical photometry and spectroscopy for nearly two billion objects \citep{gaia_mission}. The third \textit{Gaia} data release (\textit{Gaia} DR3) presented data collected during the first 34 months of the \textit{Gaia} mission \citep{2021A&A...649A...1G,gaia_DR3}. These included mean low-resolution blue and red photometer (BP/RP or XP) spectra for approximately 219 million objects (most of these brighter than $G=17.65$ mag). The \textit{Gaia} XP spectra consist of low-resolution ($R\sim30-100$, \citealp{xp_Carrasco}) spectrophotometry in the range $330-1050$ nm. Essentially, they can be understood as low-resolution aperture-free spectra averaged from many transits. The spectra are provided as an array of coefficients, needed to build the spectra from a continuous linear combination of Hermite functions. The coefficients are divided into two sets, corresponding to the two \textit{Gaia} passbands, BP and RP (\citealp{2021A&A...649A...3R}). We refer the reader to the work of \citet{xp_DeAngeli} and \citet{xp_Montegriffo} for a detailed description of the \textit{Gaia} DR3 XP spectra.

The wavelength range covered by the \textit{Gaia} RP spectra includes the H\textalpha{} line. Therefore, the \textit{Gaia} XP spectra, combined with \textit{Gaia} astrometric and photometric measurements, constitute a powerful tool for a homogeneous characterisation of the YSOs in the Galaxy. In particular, we use the information on the H\textalpha{} line obtained from the \textit{Gaia} XP spectra to identify accreting YSO candidates and derive their accretion properties. We limit the analysis to the solar neighbourhood (within 500\,pc of the Sun). This distance was chosen because it includes the most studied nearby star-forming regions, providing well-characterised samples to use for comparison and calibration, while also limiting the impact of interstellar extinction and the blending of more distant sources. 

The paper is structured as follows. In Sect. \ref{Sect:methodology}, we present the methodology used to derive accretion luminosities and mass accretion rates from the H\textalpha{} line observed in the XP spectra. This results in a table of accretion properties for 145\,975 sources within 500\,pc. We apply various filtering strategies to select specific sub-samples of YSO candidates from this table, and obtain a bona fide sample of 1\,945 YSOs. In Sect. \ref{sec:all-sky} we present an all-sky view of accretion in YSOs in the solar neighbourhood. We look for correlations between accretion properties and stellar parameters in the sub-samples of YSOs. In Sect \ref{Sect:ScoCen_section}, we apply the derived accretion properties to study the evolution of accretion with age in the Sco-Cen star-forming complex, and other star-forming regions. We summarise our conclusions in Sect. \ref{Sect:conclusions}.

\section{Methodology} \label{Sect:methodology}

\subsection{Sample selection}\label{Sect:sample_selection}

We retrieved from the \textit{Gaia} ESA Archive\footnote{\url{https://gea.esac.esa.int/archive}} all sources within 500\,pc, for which H\textalpha{} pseudo-equivalent width measurements (pEW) are available in the \textit{Gaia} DR3 table of astrophysical parameters (\texttt{astrophysical\_parameters}, \citealp{ast_par_creevy}). This table contains information on the "pseudo-equivalent width" (pEW) of the H\textalpha{} line, stored as \texttt{ew\_espels\_halpha}, together with other astrophysical parameters determined from the combination of the \textit{Gaia} mean astrometric, photometric, and both high- and low-resolution spectroscopic data. The pEWs were calculated as part of the Extended Stellar Parametrizer for Emission-Line Stars (ESP-ELS) module (\citealp{pEW_comparison}), which searches for emission lines in the H\textalpha{} wavelength range on the RP spectra for targets brighter than magnitude G=17.65\,mag. The H\textalpha{} pEWs were computed relative to a local pseudo-continuum, estimated between the flux at 646 and 670\,nm, by summing the contributions from the flux samples that fall in this range. To ensure the best completeness possible within 500\,pc, we included all sources compatible with distances $<500$\,pc according to the geometric priors of \citet{geom_dist}. The resulting dataset contains $10\;753\;718$ sources. The ADQL query used to obtain this sample is detailed in Appendix \ref{Appendix A}.

By definition, the sample of all sources with pEW contains all sources with public XP spectra, although not all sources with pEW have publicly available XP spectra in the \textit{Gaia} Archive (\citealp{xp_DeAngeli,pEW_comparison}). This is due to the fact that some XP spectra were removed from the final \textit{Gaia} data release because of a variety of reasons (see Section 4 of \citealp{xp_DeAngeli}). In particular, many YSOs were removed based on a literature search of known YSOs (private communication). Hence, most of the well-known and studied YSOs do not have public \textit{Gaia} XP spectra, but do have pEWs reported. It is for this reason that we anchor the calibrations of this work to the pEWs (Sect. \ref{sec:pEW-EWlit_correlation}), and only use the \textit{Gaia} XP spectra in our analysis when available. Studies after \textit{Gaia} data release 4 (DR4) might prefer to limit their analyses to the coefficient space of the XP spectra for population-dedicated studies like the one of this work.

For those sources with XP spectra available in their continuous representation, the \texttt{xp\_continuous\_mean\_spectrum} table provides arrays of the expansion coefficients to be applied to the basis Hermite functions and the corresponding covariance matrices, for BP and RP separately. An overview of XP spectra processing techniques and calibration models is provided in \cite{xp_Carrasco}, \cite{xp_DeAngeli}, and \cite{xp_Montegriffo}. From the continuous representation of the spectra, sampled spectra representation (i.e. in the form of integrated flux vs pixel) can be obtained using the Python package \texttt{GaiaXPy} \citep{GaiaXPy}. This software converts the basis coefficients into a spectrum sampled on a discrete wavelength grid. However, sampling the spectra results in a loss of information \citep{gaia_DR3}. \texttt{GaiaXPy} also provides a \texttt{linefinder} tool which detects lines and extrema in the spectra in its continuous representation, using the approach described in \cite{find_lines}. Hence, \texttt{linefinder} can be used to search for the H\textalpha{} emission line, and measure its properties. The line properties returned by \texttt{linefinder} include the value of flux at the wavelength corresponding to the extremum (flux\textsubscript{lf}), the difference between the line flux and the flux of the estimated continuum (depth\textsubscript{lf}), and the distance between the two closest inflection points (width\textsubscript{lf}).

We applied \texttt{linefinder} to all sources with public XP spectra within 500\,pc. \texttt{linefinder} identifies  H\textalpha{} emission for $2\;756\;293$ of them. This set has a high degree of contaminants, mostly M dwarfs which molecular bands mimic H\textalpha{} emission in very low-resolution spectra. In order to characterise what fraction of sources have real H\textalpha{} emission, we compared with the sample of known YSOs presented in \citet{Manara_TTauri}, \citet{Vioque_2018}, and \citet{Vioque_2022}. We concluded that YSOs have mostly width\textsubscript{lf}\,$<25$\,nm and depth\textsubscript{lf} $>10^{-17}$ W/nm/m\textsuperscript{2} as traced by \texttt{linefinder}. This allowed us to reduce the number of YSO-like H\textalpha{} emitters with public XP spectra within 500\,pc to $1\,106$ objects. However, as explained above, most YSOs do not have public XP spectra but do have public pEWs. To identify true H\textalpha{} emitters using the pEW alone, we calibrated the pEW measurements using the width\textsubscript{lf} and depth\textsubscript{lf} of \texttt{linefinder} for those objects with both pEW and XP data. We concluded that considering sources with pEW\,$<-1.0$\,nm is the best compromise between completeness and minimising the number of contaminants when selecting YSO H\textalpha{} emitters. This threshold reduces the sample of $10\,753\,718$ sources to $7\,232$ sources. We note, however, that a less conservative threshold for true YSO H\textalpha{} emitters is $\textrm{pEW}<-0.5$\,nm ($150\,383$ sources). By imposing a limit of $\textrm{pEW}<-0.5$\,nm, rather than pEW\,$<-1.0$\,nm, we lower the threshold in line flux and are therefore sensitive to fainter H\textalpha{} emitters, significantly increasing the sample size, at the price of increasing the contamination of the sample. Because in this work we had other means to filter true YSO sources (Fig. \ref{fig:selection_diagram}), we proceeded by considering all sources within the threshold of pEW\,$<-0.5$\,nm. 

For the sample of sources with pEW\,$<-0.5$\,nm, 130\,109 sources (or 86.5\%) have XP spectra available. For these sources, we used \texttt{linefinder} without truncation to obtain additional measurements of the H\textalpha{} line. We did not truncate the coefficients as the emission-lines are likely described by the higher order coefficients (\citealp{xp_Carrasco}, \citealp{xp_DeAngeli}).

We discarded all sources with \textit{Gaia} XP spectra that have \texttt{linefinder}-measured width$_{lf}>25$ nm, as we found \texttt{linefinder} H\textalpha{} detections for sources in this range are mostly M-dwarf contaminants. This is due to the fact that the broad titanium oxide molecular bands of M-dwarfs are often misclassified by \texttt{linefinder} as H\textalpha{} emission (see Appendix \ref{Appendix B}). This excluded $4\,408$ sources from the table, leaving a dataset of 145\,975 sources. We kept sources for which \texttt{linefinder} does not detect a H\textalpha{} emission as this means that they do not have the broad features typical of M-dwarfs. We also kept all $20\,274$ sources with no XP spectra publicly available. However, one of the criteria for selecting which sources have public XP spectra is based on the number of \textit{Gaia} CCD transits. Hence, due to the \textit{Gaia} scanning law sources without XP spectra are not-uniformly distributed in the sky \citep{xp_DeAngeli}. In Sect. \ref{Sect:YSO_flags_criteria} we filter sources in regions affected by this to produce a homogeneous distribution of YSOs.

The remaining sample of $145\,975$ sources is the sample with which we proceed for the rest of this work (Table \ref{table:AccretionTable_extract}) unless stated otherwise. A more thorough description of the analyses described in this paragraph is presented in Appendix \ref{Appendix B}.

\subsection{Derivation of H-alpha equivalent widths from ESP-ELS pseudo-equivalent widths} \label{sec:pEW-EWlit_correlation}

In this section we evaluate how the pseudo-equivalent widths reported in \citet{pEW_comparison} from the \textit{Gaia Astrophysical Parameters Inference System} (\textit{Apsis}, \citealp{ast_par_creevy}), compare to observed equivalent widths in YSOs obtained with higher-resolution spectrographs. We compiled a sample of YSOs with literature values of H\textalpha{} equivalent width ($\text{EW}_{\text{H\textalpha}}$) obtained from VLT/X-Shooter observations ($R\sim10\,000$) and \textit{Gaia} pEW. The combined sample contains T Tauri stars from the following star-forming regions: Lupus (41 sources, \citealp{Alcala_lupus1}; \citealp{Alcala_lupus2}), Chamaeleon (12 sources, \citealp{2016A&A...585A.136M}), and Upper Scorpius (27 sources, \citealp{2020A&A...639A..58M}). We also included 58 Herbig Ae/Be stars \citep{Herbigs_bonafide} to consider a more ample population and derive a correlation that is valid for a broader mass range. We note that these observed $\text{EW}_{\text{H\textalpha}}$ are not corrected from the underlying line absorption. We looked for a correlation between the \textit{Gaia} pEW and literature $\text{EW}_{\text{H\textalpha}}$ values, of the form
\begin{equation}
    \log(-\text{EW}_{\text{H\textalpha}}) = m \cdot \log(-\text{pEW}) + c + \epsilon_1,
    \label{eqn:EWlit_pEW_relation}
\end{equation}
where $m$ and $c$ are the regression coefficients, and $\epsilon_1$ is the intrinsic random scatter about the regression, assumed to be normally distributed with mean zero and error $\sigma_{\epsilon_1}$. The parameters were found by fitting the correlation using the Bayesian method to account for measurement errors described in \citet{linmix}\footnote{linmix Python package \url{https://linmix.readthedocs.io/en/latest/}}. We considered the reported uncertainties in both $\text{EW}_{\text{H\textalpha}}$ and pEW. Considering all sources with emission gives a Spearman correlation coefficient of 0.89, but we found a correlation coefficient of only 0.6 for the sources with values of pEW\,$> -0.1$ or $\text{EW}_{\text{H\textalpha}} > -0.1$. Therefore, we discarded the 18 points in this limit. The remaining sample contains 120 sources, with a Spearman correlation coefficient of 0.88. By fitting to the latter sample, the following parameters were obtained:
\begin{equation}
\begin{aligned}
    m &= 1.13 \pm 0.04 \\
    c &= 0.41 \pm 0.02\\
    \sigma&_{\epsilon_1} = 0.18 .
\end{aligned}
\label{eqn:EWlit_pEW_coefficients}
\end{equation}

\begin{figure}
    \centering
    \includegraphics[width=\linewidth]{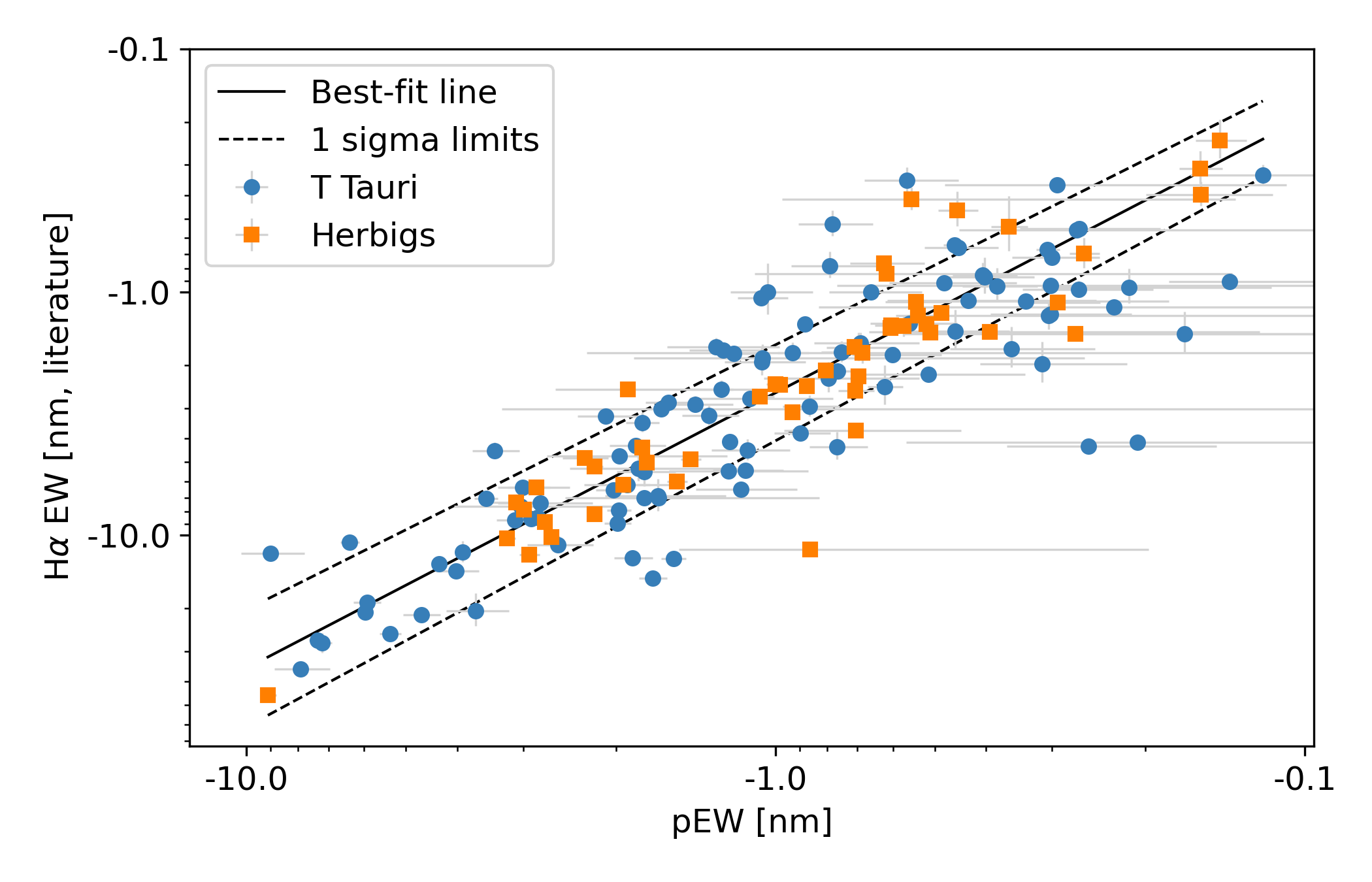}
    \caption{Correlation between H\textalpha{} equivalent widths from medium-resolution spectra and \textit{Gaia} H\textalpha{} pseudo-equivalent widths, for a sample of YSOs (T Tauri and Herbig stars). The black lines represent the best linear fit to the data and its uncertainties, $\log(-\text{EW}_{\text{H\textalpha}}) = (1.13 \pm 0.04) \cdot \log(-\text{pEW}) + (0.41 \pm 0.02) \pm 0.18$.}
    \label{fig:EWlit_pEW_relation}
\end{figure}

The correlation is shown in Fig. \ref{fig:EWlit_pEW_relation}. We used this correlation between pEW and $\text{EW}_{\text{H\textalpha}}$ to derive EW$_{\text{H\textalpha}}$ for all stars in our dataset with pEW\,$<-0.5$\,nm (Sect. \ref{Sect:sample_selection}, Appendix \ref{Appendix B}). The errors in $m$ and $c$, the intrinsic scatter $\sigma_{\epsilon_1}$ (Eq. \ref{eqn:EWlit_pEW_coefficients}) and the pEW uncertainties were accounted for and propagated through bootstrapping with 1000 samples. We note that for the 1.1\% of sources where the reported fractional error in pEW is greater than 100\%, we set it to 90\% instead. Each reported value of  $\text{EW}_{\text{H\textalpha}}$ and its lower and upper limit are given, respectively, by the median and 16\textsuperscript{th} and 84\textsuperscript{th} percentile. Measured $\text{EW}_{\text{H\textalpha}}$ are reported in Table \ref{table:AccretionTable_extract}.

\subsection{Derivation of the H-alpha line flux} \label{sec:F_Halpha_derivation}

To convert EW$_{\text{H\textalpha}}$ into a H\textalpha{} line flux, it is necessary to know the continuum flux density at the H\textalpha{} line ($F_{\text{cont}}$). For the YSOs with XP spectra available, $F_{\text{cont}}$ can be obtained from the spectra using \texttt{linefinder} \citet{find_lines}. We combined the YSOs from the Sco-Cen star-forming region identified in \citet{ScoCen_SigMA_sources} that have have pEW\,$<-0.5$\,nm with the T Tauri stars from \citet{Manara_TTauri} and the Herbig stars from \citet{Herbigs_bonafide}. We selected the sources with XP spectra, ran \texttt{linefinder} without truncation on their spectra and then selected the sources for which \texttt{linefinder} reports values with depth$_{lf}>10^{-17}$ W/nm/m$^2$ and width$_{lf}<25$ nm. We selected these values to reduce the number of potentially misclassified H\textalpha{} emitters (see Appendix \ref{Appendix B}). The resulting sample contains 114 sources.

We then calculated $F_{\text{cont}} = \text{flux}_{lf}-\text{depth}_{lf}$. We obtained uncertainty estimates for $F_{\text{cont}}$ by setting the errors equal to the difference between the values of $F_{\text{cont}}$ calculated using \texttt{linefinder} when truncating the coefficients which represent the spectra to a recommended value and when not truncating them. We also set the errors to a minimum of $10\%$ as a conservative estimate. We then fit a relation between $F_{\text{cont}}$ and the \textit{Gaia} DR3 mean magnitude in the integrated RP band (RP), assuming an error of $5\%$ for RP. We fit a relation of the form
\begin{equation}
    \log(F_{\text{cont}}) = a \cdot \text{RP} + b + \epsilon_2 ,
    \label{eqn:F_cont_RP}
\end{equation}

where $a$ and $b$ are the regression coefficients, and $\epsilon_2$ is used to account for the intrinsic random scatter. Using a Bayesian method to account for errors \citep{linmix}, we obtained the values of $a$ and $b$, with errors, and of the error from the intrinsic scatter, $\sigma_{\epsilon_2}$ (see Table \ref{table:Fcont_RP_coefficients}). The plot of the correlation is shown in Appendix \ref{Appendix:Fcont_RP_correlation}. We used this relation to derive $F_{\text{cont}}$ from \textit{Gaia} RP for all sources in our original dataset with pEW\,$<-0.5$\,nm. Then, we derived the H\textalpha{} line flux as $F_{\text{H\textalpha}} = \text{EW}_{\text{H\textalpha}} \cdot F_{\text{cont}}$. The uncertainties on the fit parameters and RP and the intrinsic scatter were propagated consistently through bootstrapping. The median value as well as the 16\textsuperscript{th} and 84\textsuperscript{th} percentiles were calculated for both $F_{\text{cont}}$ and $F_{\text{H\textalpha}}$, and they are presented in Table \ref{table:AccretionTable_extract}.

\subsection{Accounting for interstellar extinction}\label{ext_section}

In the approach presented in Sect. \ref{sec:F_Halpha_derivation} for deriving the H\textalpha{} line flux $F_{\text{H\textalpha}}$, interstellar extinction is not taken into account. We repeated the derivation of $F_{\text{H\textalpha}}$ two more times. First, we accounted for extinction using the broadband extinction provided as part of \textit{Gaia} DR3. The \textit{Gaia} DR3 \texttt{astrophysical\_parameters} table provides the broadband extinction in the \textit{Gaia} passbands inferred by the \textit{General Stellar Parametrizer from Photometry Aeneas} using BP/RP spectra, apparent G magnitude and parallax \citep[GSP-Phot extinction]{extinction_GSPPHOT}. 

However, only 68\% of the sample of $145\,975$ sources with pEW$<-0.5$\,nm has GSP-Phot extinctions. For this reason, we also estimated interstellar extinctions by taking the median value of the GSP-Phot extinctions of neighbouring stars (`med-GSP-Phot extinction'). We did this for 99\% of the sample, which have at least one other source within a 5 pc radius sphere (94\% of them have five or more sources in this volume).

We calculated the extinction-corrected RP magnitude as $\text{RP}_{\text{corrected}} = \text{RP} - A_{RP}$, where $A_{RP}$ is the extinction in the corresponding band. Upper and lower limits for $\text{RP}_{\text{corrected}}$ were obtained by considering the uncertainty in $A_{RP}$. This uncertainty is provided by \textit{Gaia} DR3 in the case of GSP-Phot extinctions, and we took the mean of the 25\textsuperscript{th} and 75\textsuperscript{th} percentile errors as the uncertainty in med-GSP-Phot extinctions. We obtained the extinction-corrected continuum flux from the equation: 
\begin{equation}
    F_{\text{cont,corrected}} = F_{\text{cont}} \cdot 10^{-0.4 A_{RP}}.
    \label{F_cont_corr}
\end{equation}
We repeated the fit from Sect. \ref{sec:F_Halpha_derivation} (Eq. \ref{eqn:F_cont_RP}), using values of $\text{RP}_{\text{corrected}}$ and $F_{\text{cont,corrected}}$, extinction-corrected with GSP-Phot and with med-GSP-Phot extinctions. The values of the parameters relating the continuum flux density at the H\textalpha{} line with the RP magnitude under these extinction corrections are reported in Table \ref{table:Fcont_RP_coefficients}, and the plot of the relation between the two quantities can be found in Appendix \ref{Appendix:Fcont_RP_correlation}. We then obtained new values of $F_{\text{H\textalpha}}$ for all sources with pEW\,$<-0.5$, using the two extinction estimations, propagating errors through bootstrapping (Table \ref{table:AccretionTable_extract}).

\begin{table}
\caption[]{Coefficients of linear fit described in Equation \ref{eqn:F_cont_RP}, $\log(F_{\text{cont}}) = a \cdot \text{RP} + b + \epsilon_2$, which describes the correlation between H\textalpha{} continuum flux density and RP magnitude, for three extinction correction regimes.}
\label{table:Fcont_RP_coefficients}
 $$
\begin{array}{p{0.3\linewidth}cccc}
    \hline\hline
    \noalign{\smallskip}
         Extinction &  a  &  b & \sigma_{\epsilon_2}\\
        \noalign{\smallskip}
        \hline
        \noalign{\smallskip}
        No correction & -0.48 \pm 0.01 & -10.2 \pm 0.1 & 0.06\\
        GSP-Phot      & -0.48 \pm 0.01 & -10.2 \pm 0.1 & 0.05\\
        med-GSP-Phot      & -0.49 \pm 0.01 & -10.2 \pm 0.1 & 0.05\\
        \noalign{\smallskip}
    \hline  
\end{array}
     $$ 
\end{table}

\begin{figure*}
    \centering
    \includegraphics[width=1\linewidth]{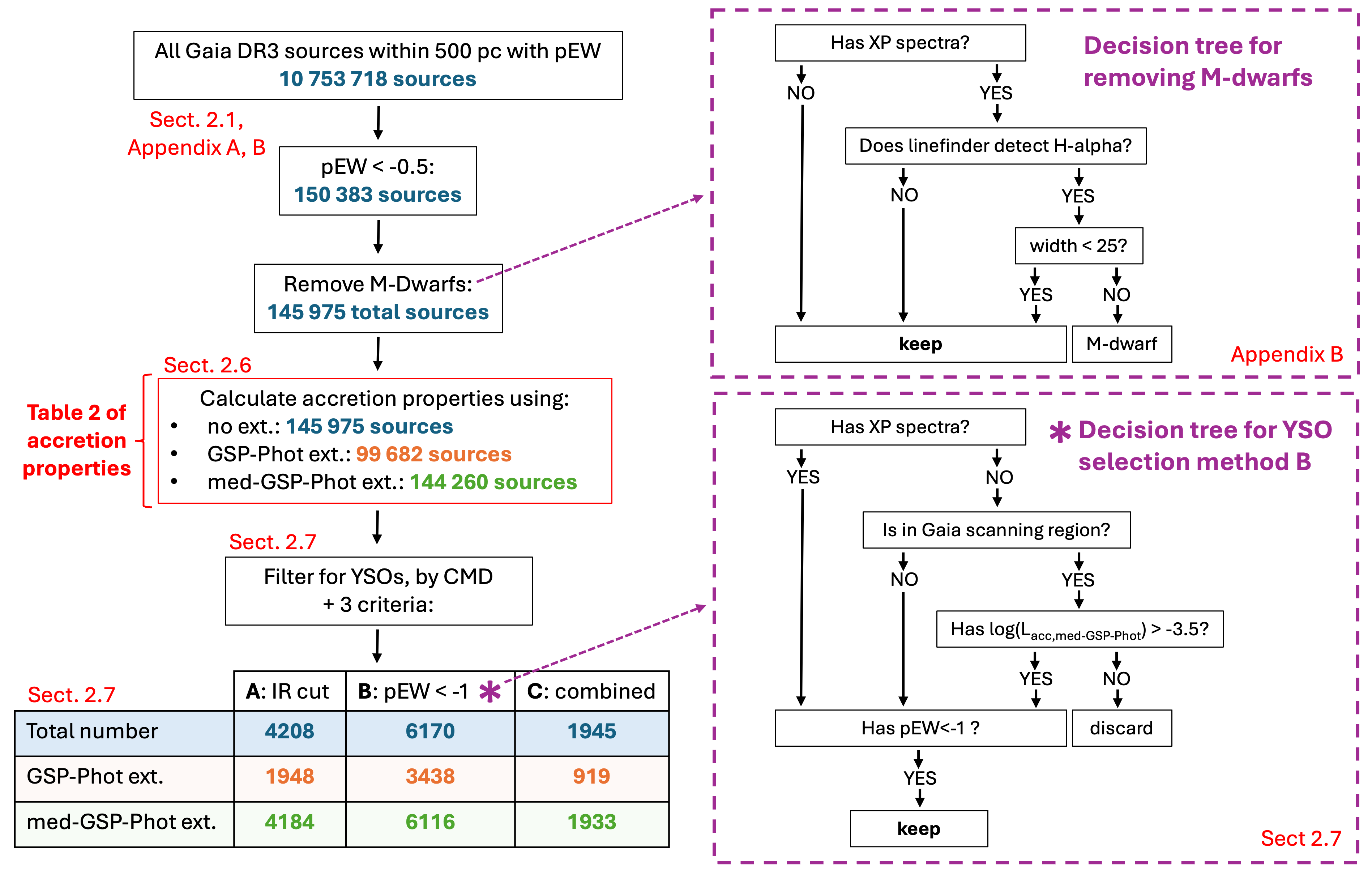}
    \caption{Diagram illustrating the various selections applied to the original sample of  all \textit{Gaia} DR3 sources within 500\,pc with measured pEW to obtain the final sample of YSO candidates (Table \ref{table:AccretionTable_extract}), and the different quality flags that can be applied to further refine the purity of the sample. The two boxes on the right show two of the steps in more detail.}
    \label{fig:selection_diagram}
\end{figure*}

\subsection{Derivation of stellar parameters} \label{sec:derivation_stellar_parameters}

We estimated stellar parameters for the sample of $145\,975$ stars within 500\,pc and with $\text{pEW}<-0.5$\,nm (stellar masses, effective temperature, stellar luminosities, stellar radii, and surface gravities). For this, we used the extinction estimations of the previous section. We corrected the \textit{Gaia} G\textsubscript{BP}-G\textsubscript{RP} colours from extinction, and derived absolute M\textsubscript{G} magnitudes by using the geometric distances of \citet{geom_dist}. We then placed our sources in the dereddened colour-magnitude diagram (CMD), from where we derived stellar parameters using theoretical evolutionary tracks. We used the \citet{BHAC15} tracks which are optimised for pre-main sequence and main sequence low-mass stars (extending to 1.4\,M$_\odot$). Stellar parameters were derived assuming both GSP-Phot extinction and med-GSP-Phot extinction. We set the errors in the colour-magnitude diagram to be at minimum 0.1\,mag in colour and in absolute magnitude. This decision is due to the fact that \textit{Gaia} uncertainties are sometimes underestimated, producing unrealistically accurate positions in the CMD. Measured stellar parameters for both sets of extinction corrections are presented in Table \ref{table:AccretionTable_extract}. Sources outside of the \citet{BHAC15} tracks in the CMD have no stellar parameters associated. Therefore, stellar parameters calculated from GSP-Phot extinction are available for a total of 80\,348 stars, and stellar parameters calculated from med-GSP-Phot extinction are available for a total of 140\,083 stars.

\subsection{Accretion luminosities and mass accretion rates} \label{sec:Lacc_derivation}

The H\textalpha{} line luminosity can be obtained as $L_{\text{H\textalpha}} = 4\pi d^2 \cdot F_{\text{H\textalpha}}$, where $d$ is the geometric distance from \citet{geom_dist}. The accretion luminosity can be derived as:
\begin{equation}
    \log(L_{\text{acc}}/L_{\odot}) = A + B \cdot \log(L_{\text{H\textalpha}}/L_{\odot}),
    \label{eqn:L_acc}
\end{equation}

where $A$ and $B$ are constants. We used $A = 1.13 \pm 0.05$ and $B = 1.74 \pm 0.19$, which are the values determined in \cite{Alcala_lupus2} for classical T Tauri stars. Using this equation and the results of Sects. \ref{sec:F_Halpha_derivation} and \ref{ext_section}, we derived accretion luminosities for all sources in our sample and for the three extinction regimes described in Sect. \ref{ext_section} (no extinction, GSP-Phot extinction, and med-GSP-Phot extinction). 

Following the approach presented in \citet{Alcala_lupus2}, we converted the accretion luminosities into mass accretion rates, $\dot{M}_{\text{acc}}$, using the relation
\begin{equation}
    \dot{M}_{\text{acc}} = \left( 1 - \frac{R_\star}{R_{\text{in}}} \right) ^{-1} \frac{L_\text{acc}R_\star}{GM_\star} \approx 1.25 \frac{L_\text{acc}R_\star}{GM_\star},\label{eqn:m_acc}
\end{equation}

where $R_\star$ and $M_\star$ are the stellar radius and mass, respectively, and $R_{\text{in}}$ is the inner-disc radius, which as in \citet{Alcala_lupus2} we assumed to be $R_{\text{in}}=5R_\star$ for comparison with previous results. We calculated mass accretion rates for all sources with pEW\,$< -0.5$\,nm, correcting for extinction with both GSP-Phot extinction and med-GSP-Phot extinction. We note that equations \ref{eqn:L_acc} and \ref{eqn:m_acc} assume magnetospheric accretion and that the H\textalpha{} emission component is tracing the accretion flows (\citealp{2016ARA&A..54..135H}).

All errors, including those of the distance, of constants $A$ and $B$, of the stellar parameters, and of all initial quantities and fit parameters, as well as the intrinsic scatter in the relationships, discussed in the previous sections were propagated consistently to the final values of $L$\textsubscript{acc} and $\dot{M}_{\text{acc}}$ through bootstrapping with 1000 steps. For each value we report the median and the 16\textsuperscript{th} and 84\textsuperscript{th} percentiles.

In summary, we produced a table of accretion luminosities and mass accretion rates for all 145\,975 sources within 500\,pc with pEW\,$<-0.5$\,nm, excluding the sources with \texttt{linefinder}-measured width$_{lf}>25$ nm. An extract of the table is presented in Table \ref{table:AccretionTable_extract}, and the full table is available as online material.

\begin{table*}
\caption{Parameters derived in this work from \textit{Gaia} XP spectra and pEW for H\textalpha{} emitters within 500\,pc ($145\,975$ sources).}
\label{table:AccretionTable_extract}      
\centering 
\resizebox{0.9\textwidth}{!}{\begin{tabular}{c c c c c c}
\hline\hline\\[-0.9em]   
 & & \multicolumn{4}{c}{GSP-Phot extinction}\\
    \cline{3-6}  
    \\[-0.8em] 
Gaia DR3 Source ID & pEW [nm] & Mass [$M_\odot$] & log(Luminosity/$L_\odot$) & log($L_{\text{acc}}/L_\odot$)  & $\dot{M}_\text{acc}$ [$M_\odot$ yr$^{-1}$] \\
\noalign{\smallskip}
\hline                       
\noalign{\smallskip}
2130017669575110144 & $-0.81 \pm 0.21$ & $0.052 \pm 0.003$ & $-2.82 \pm 0.14$ & $-5.59$ ($-6.09$, $-5.10$) & $3.48\mathrm{E}{-13}$ ($1.13\mathrm{E}{-13}$, $1.10\mathrm{E}{-12}$) \\
1287845363887270400 & $-1.26 \pm 0.06$ & $0.054 \pm 0.004$ & $-2.22 \pm 0.11$ & $-4.52$ ($-4.96$, $-4.04$) & $7.31\mathrm{E}{-12}$ ($2.62\mathrm{E}{-12}$, $2.17\mathrm{E}{-11}$) \\
5989145703171514240 & $-1.83 \pm 0.24$ & $0.118 \pm 0.006$ & $-2.10 \pm 0.10$ & $-4.01$ ($-4.45$, $-3.60$) & $1.01\mathrm{E}{-11}$ ($3.86\mathrm{E}{-12}$, $2.65\mathrm{E}{-11}$) \\
3438864281072161408 & $-0.54 \pm 0.20$ & $0.121 \pm 0.006$ & $-2.82 \pm 0.14$ & $-5.75$ ($-6.34$, $-5.24$) & $8.81\mathrm{E}{-14}$ ($2.34\mathrm{E}{-14}$, $2.86\mathrm{E}{-13}$) \\
4054163666720801408 & $-0.52 \pm 0.13$ & $0.123 \pm 0.006$ & $-2.65 \pm 0.13$ & $-5.43$ ($-5.92$, $-4.92$) & $2.04\mathrm{E}{-13}$ ($6.54\mathrm{E}{-14}$, $6.74\mathrm{E}{-13}$) \\
2717368255618034560 & $-0.71 \pm 0.32$ & $0.111 \pm 0.006$ & $-2.91 \pm 0.15$ & $-5.82$ ($-6.44$, $-5.25$) & $7.47\mathrm{E}{-14}$ ($1.83\mathrm{E}{-14}$, $2.81\mathrm{E}{-13}$) \\
5343078168255320448 & $-1.33 \pm 0.32$ & $0.163 \pm 0.008$ & $-1.86 \pm 0.09$ & $-3.95$ ($-4.39$, $-3.50$) & $1.06\mathrm{E}{-11}$ ($3.85\mathrm{E}{-12}$, $3.09\mathrm{E}{-11}$) \\
5842032249951369088 & $-0.99 \pm 0.37$ & $0.087 \pm 0.005$ & $-2.26 \pm 0.11$ & $-4.83$ ($-5.39$, $-4.28$) & $1.88\mathrm{E}{-12}$ ($4.90\mathrm{E}{-13}$, $6.50\mathrm{E}{-12}$) \\
1721746737067096832 & $-1.12 \pm 0.27$ & $0.139 \pm 0.012$ & $-2.46 \pm 0.12$ & $-4.89$ ($-5.42$, $-4.43$) & $7.70\mathrm{E}{-13}$ ($2.34\mathrm{E}{-13}$, $2.18\mathrm{E}{-12}$) \\
5788551695132821376 & $-4.78 \pm 0.22$ & $0.094 \pm 0.005$ & $-2.05 \pm 0.10$ & $-3.51$ ($-3.93$, $-3.07$) & $4.61\mathrm{E}{-11}$ ($1.70\mathrm{E}{-11}$, $1.24\mathrm{E}{-10}$) \\
...	& ... & ... & ...	&...	&...\\
\hline
\end{tabular}}
\\[+0.8em] 
\resizebox{\textwidth}{!}{\begin{tabular}{ccccccccc}
\hline\hline\\[-0.9em]   
\multicolumn{4}{c}{med-GSP-Phot extinction} & & \multicolumn{4}{c}{YSO flags}\\
    \cline{1-4}  
    \cline{6-9}
    \\[-0.8em] 
Mass [$M_\odot$] & log(Luminosity/$L_\odot$) & log($L_{\text{acc}}/L_\odot$)  & $\dot{M}_\text{acc}$ [$M_\odot$ yr$^{-1}$] & & flag\_CMD & A. flag\_IR & B. flag\_pEW & C. flag\_combined \\
\hline
\noalign{\smallskip}
$0.052 \pm 0.003$ & $-2.82 \pm 0.14$ & $-5.56$ ($-6.11$, $-5.05$) & $3.74\mathrm{E}{-13}$ ($1.07\mathrm{E}{-13}$, $1.25\mathrm{E}{-12}$) & & true & false & false & false \\
$0.057 \pm 0.003$ & $-2.23 \pm 0.11$ & $-4.51$ ($-4.96$, $-4.08$) & $6.64\mathrm{E}{-12}$ ($2.44\mathrm{E}{-12}$, $1.85\mathrm{E}{-11}$) & & true & false & true & false \\
$0.118 \pm 0.006$ & $-2.10 \pm 0.10$ & $-3.99$ ($-4.44$, $-3.56$) & $1.04\mathrm{E}{-11}$ ($3.78\mathrm{E}{-12}$, $2.83\mathrm{E}{-11}$) & & true & true & true & true \\
$0.121 \pm 0.006$ & $-2.82 \pm 0.14$ & $-5.75$ ($-6.32$, $-5.21$) & $8.79\mathrm{E}{-14}$ ($2.36\mathrm{E}{-14}$, $2.99\mathrm{E}{-13}$) & & true & false & false & false \\
$0.123 \pm 0.006$ & $-2.65 \pm 0.13$ & $-5.40$ ($-5.93$, $-4.96$) & $2.22\mathrm{E}{-13}$ ($6.55\mathrm{E}{-14}$, $6.17\mathrm{E}{-13}$) & & true & false & false & false \\
$0.111 \pm 0.006$ & $-2.91 \pm 0.15$ & $-5.84$ ($-6.45$, $-5.33$) & $7.14\mathrm{E}{-14}$ ($1.80\mathrm{E}{-14}$, $2.32\mathrm{E}{-13}$) & & true & false & false & false \\
$0.054 \pm 0.035$ & $-2.24 \pm 0.19$ & $-4.76$ ($-5.27$, $-4.26$) & $4.01\mathrm{E}{-12}$ ($7.82\mathrm{E}{-13}$, $1.71\mathrm{E}{-11}$) & & true & false & true & false \\
$0.088 \pm 0.025$ & $-2.26 \pm 0.11$ & $-4.79$ ($-5.35$, $-4.29$) & $2.01\mathrm{E}{-12}$ ($6.17\mathrm{E}{-13}$, $7.11\mathrm{E}{-12}$) & & true & true & false & false \\
$0.112 \pm 0.029$ & $-2.49 \pm 0.12$ & $-5.02$ ($-5.51$, $-4.57$) & $7.42\mathrm{E}{-13}$ ($2.29\mathrm{E}{-13}$, $2.28\mathrm{E}{-12}$) & & true & false & true & false \\
$0.105 \pm 0.031$ & $-1.98 \pm 0.10$ & $-3.41$ ($-3.89$, $-2.98$) & $5.83\mathrm{E}{-11}$ ($1.87\mathrm{E}{-11}$, $1.66\mathrm{E}{-10}$) & & true & true & true & true \\
...	& ... & ... & ... & &... &... &... &...\\
\hline
\end{tabular}}

\tablefoot{Stellar parameters, accretion luminosities, and mass accretion rates are reported assuming both GSP-Phot and med-GSP-Phot extinction. While not shown here, the full table also contains \texttt{linefinder} resulting parameters, extinctions, $\text{EW}_{\text{H\textalpha}}$, and H\textalpha{} continuum flux densities, as well as supplementary data from \textit{Gaia} DR3 and allWISE. YSO quality flags are described in Sect. \ref{Sect:YSO_flags_criteria}. The full table is available at the CDS.}
\end{table*}

\subsection{Criteria for identifying YSOs} \label{Sect:YSO_flags_criteria}

Various filters can be considered to refine Table \ref{table:AccretionTable_extract} and select purer samples of YSO candidates. However, each filter introduces a noticeable selection bias towards a certain population of YSOs.  Table \ref{table:AccretionTable_extract} (Sects. \ref{Sect:sample_selection} to \ref{sec:Lacc_derivation} and Appendix \ref{Appendix B}) is intended to be the most general compilation of YSO H\textalpha{} emitter candidates within 500\,pc that can be gathered from \textit{Gaia} XP spectra. We added different `quality flags' to Table \ref{table:AccretionTable_extract} that users might want to apply, or not, depending on each individual science case (see e.g. Sects. \ref{Sec:Accretion_all_sky} and \ref{Sect:ScoCen_section}). A graphic representation of the full decision tree used to build Table \ref{table:AccretionTable_extract} and its different quality flags is presented in Fig. \ref{fig:selection_diagram}. We dedicate this section to explain these quality flags.

\begin{figure*}[ht]
    \centering
    \begin{subfigure}[b]{0.49\textwidth}
        \centering
        \includegraphics[width=\textwidth, trim=0 20 0 20, clip]{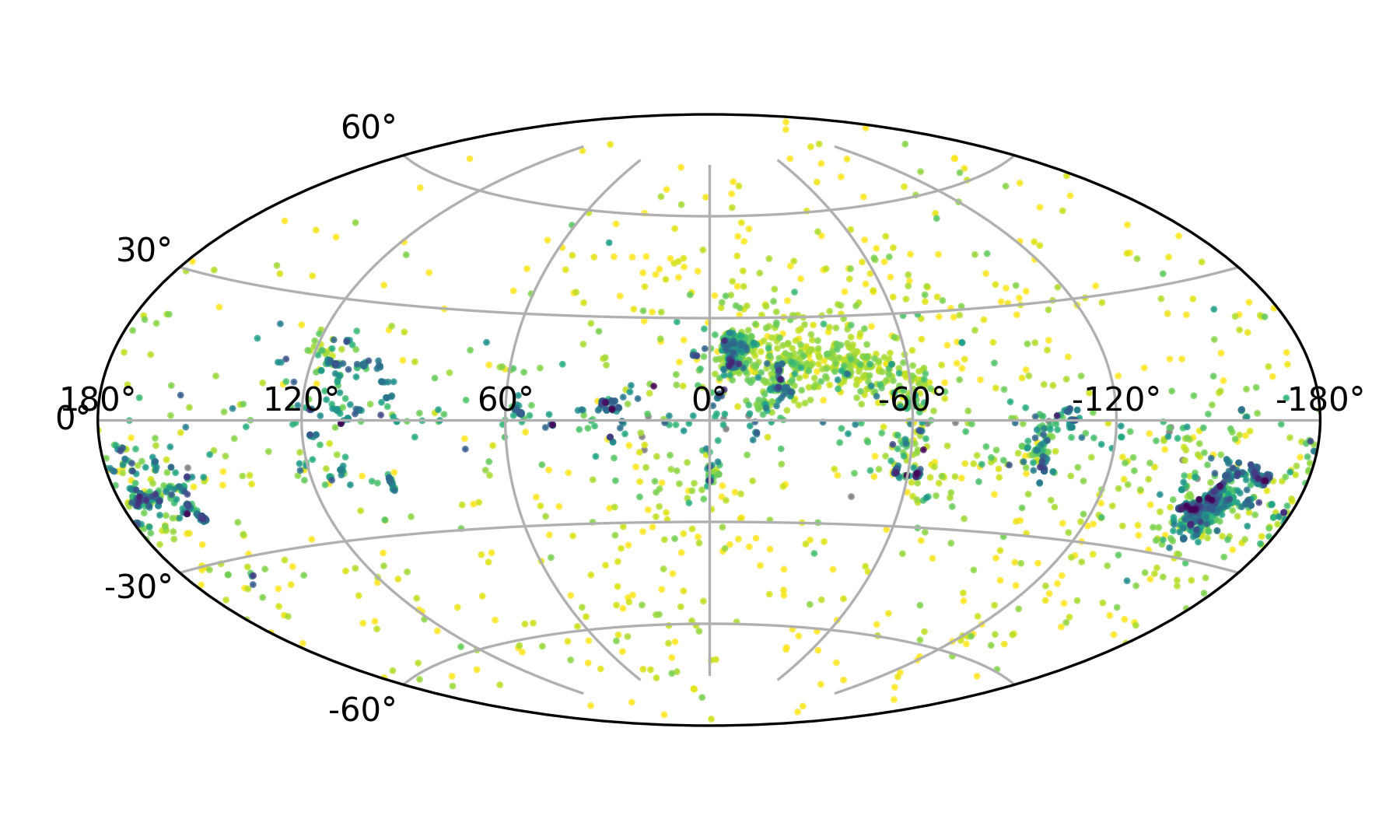} 
        \caption{}
        \label{fig:subfig1}
    \end{subfigure}
    \begin{subfigure}[b]{0.49\textwidth}
        \centering
        \includegraphics[width=\textwidth, trim=0 20 0 20, clip]{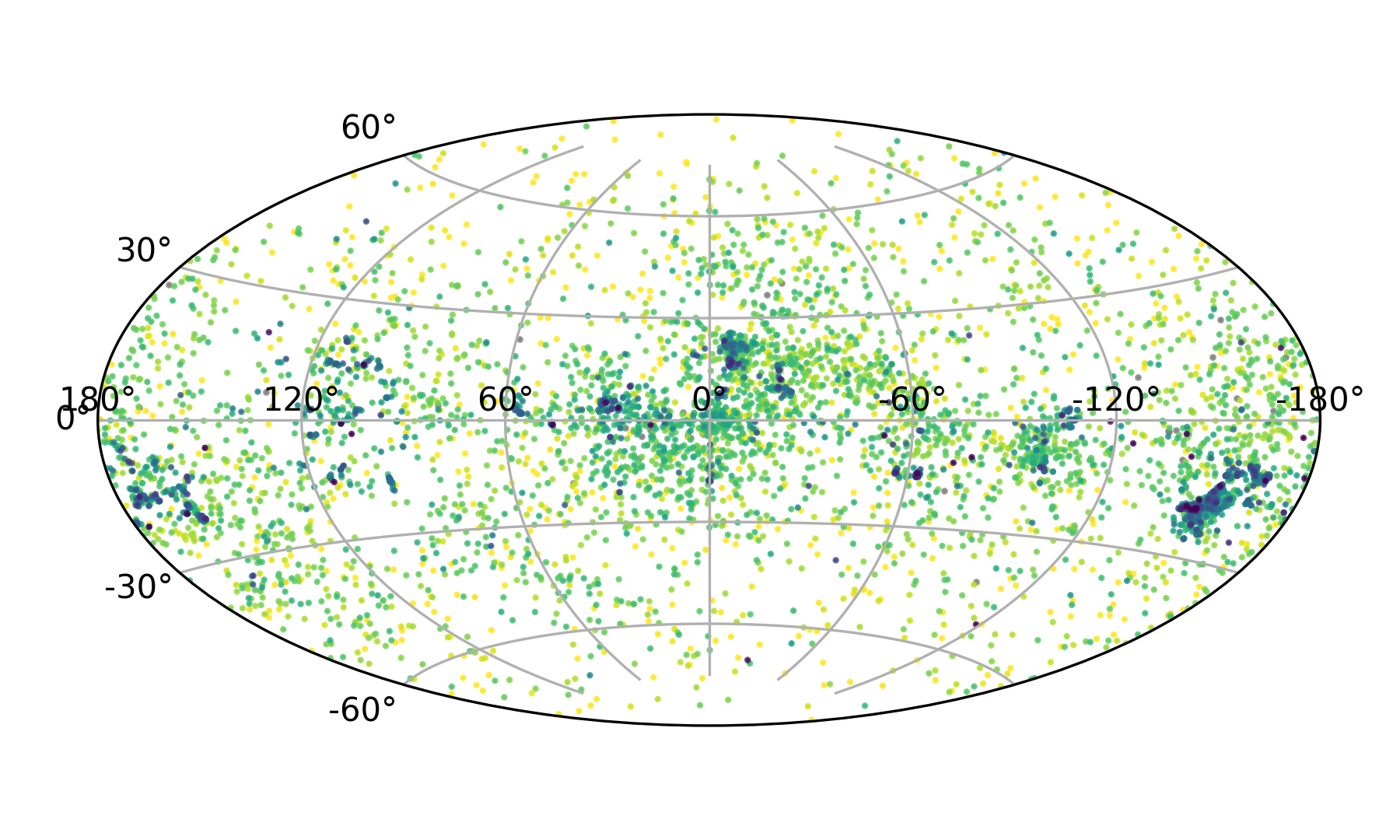} 
        \caption{}
        \label{fig:subfig2}
    \end{subfigure}

    \begin{subfigure}[b]{\textwidth}
        \centering
        \includegraphics[width=\textwidth]{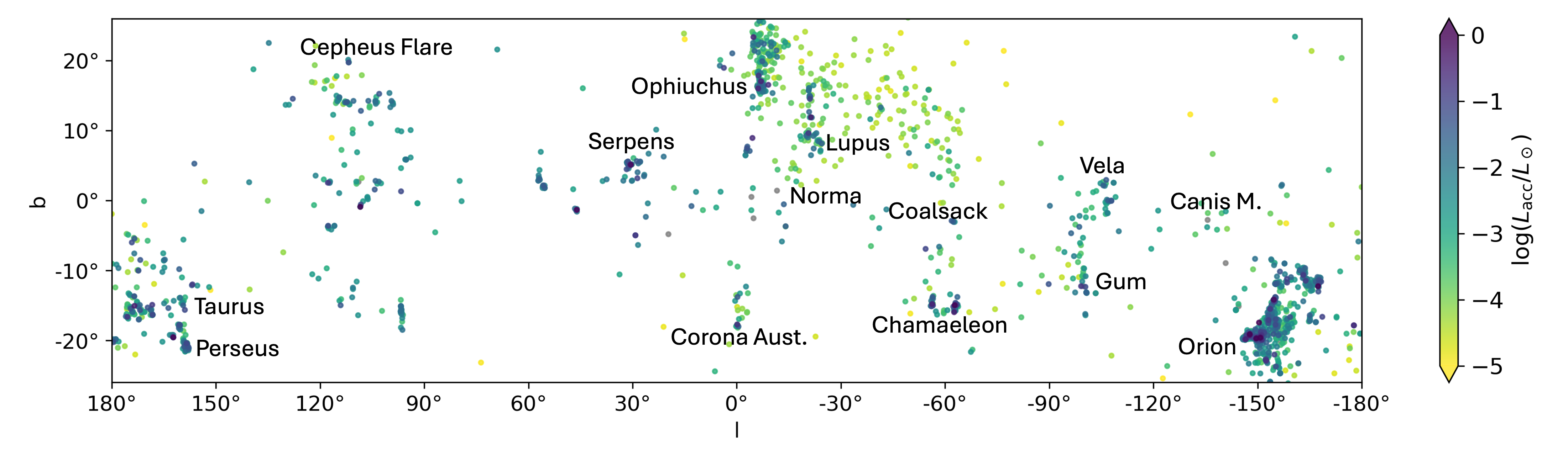} 
        \caption{}
        \label{fig:subfig3}
    \end{subfigure}
    
    \caption{Fig. (a): Sky plot in galactic co-ordinates of YSO H\textalpha{} emitters selected using criterion A (filtering by IR excess, Sect. \ref{Sect:YSO_flags_criteria}). Fig. (b): Sky plot in galactic co-ordinates of YSO H\textalpha{} emitters selected using criterion B (filtering by pEW). Fig. (c): Sky plot in galactic co-ordinates of YSO H\textalpha{} emitters selected using criterion C, which is the intersection of sources in A and B. In Fig.(c), the plot is zoomed in around the Galactic plane. The star-forming regions are labelled following \citet{SF_comparison_map}. The sources are colour-coded by their accretion luminosity calculated using med-GSP-Phot extinction (Sect. \ref{ext_section}). The few sources without med-GSP-Phot extinction are shown in grey.}
    \label{fig:SKY-plots}
\end{figure*}

First, we used \textit{Gaia} photometry to remove sources outside of the YSO locus in the colour-magnitude diagram (Fig. \ref{fig:HR_cut}). The YSO locus we defined is reported in Appendix \ref{Appendix C}. This removed 1\,494 sources from the sample that are incompatible with a YSO location in the CMD. Sources within the YSO-compatible region are flagged in Table \ref{table:AccretionTable_extract} as `flag\_CMD'. To compare with other YSO catalogues, $\sim35\,000$ sources of the resulting sample with  `flag\_CMD' are contained in the \citet{2019MNRAS.487.2522M} catalogue, and $\sim25\%$ of the \citet{2023A&A...674A..21M} sample of variable YSO candidates is contained in our sample with `flag\_CMD' (when limited to 500\,pc and considering only sources with XP spectra). To further identify YSOs in the `flag\_CMD' sample, we propose three different selection criteria:

\begin{enumerate}
    \setlength{\itemsep}{1mm}
    \item[{A}] Filter by infrared (IR) excess signalling the presence of protoplanetary discs: We performed a 2 arcsecond cross-match of \textit{Gaia} DR3 co-ordinates with AllWISE (\citealp{2014yCat.2328....0C}), which contains both 2MASS and WISE photometries. We selected the sources that show an IR excess similar to the one produced by protoplanetary discs in known YSOs. In particular, we selected sources that have $\text{J}-\text{K\textsubscript{s}} > 0.60$ mag and $\text{W1}-\text{W2} > 0.25$ mag (excluding upper limits), as is shown in Fig. \ref{fig:col_col_cut}. With this selection, we retained 95\% of the YSOs from \citet{Manara_TTauri}. This IR cut produces an all-sky sample of $4\,208$ sources (Fig. \ref{fig:SKY-plots}). We did not use W3 and W4 because their lower angular resolution often results in source blending in crowded regions, producing false positives. We note that applying this filter biases the selection towards sources with a significant amount of re-radiated emission from their protoplanetary discs, and against sources with discs which produce weak or no excess in the near IR (e.g. sources with large cavities). Indeed, we observe high accretion luminosities for sources in the IR excess regime selected by this filter (Fig. \ref{fig:col_col_cut}.), independently suggesting that the selection is biased towards more massive discs (\citealp{2016A&A...591L...3M}). Sources that satisfy this filter have the `flag\_IR' quality flag in Table \ref{table:AccretionTable_extract}.

    \item[{B}] Filter by pEW\,$<-1$\,nm: In Sect. \ref{Sect:sample_selection} (see also Appendix \ref{Appendix B}) we find pEW\,$<-1$\,nm is the pEW threshold which achieves the best balance between number of sources and reliable YSO H\textalpha{} emitters. We note that applying this filter biases the selection towards sources with stronger H\textalpha{} emission. For this quality flag, we also introduced an additional criterion to account for the non-homogeneous distribution in the sky of public XP spectra due to the \textit{Gaia} scanning law (Sect. \ref{sec:Lacc_derivation}). For the sources without XP spectra, we manually selected those that fall in the regions with poor \textit{Gaia} coverage and only keep those that have log($L_\text{acc,med-GSP-Phot extinction}/L_\odot) > -3.5$. This threshold is discussed in more detail in Appendix \ref{Appendix D}. This filter produces an all-sky sample of 6\,170 sources (Fig. \ref{fig:SKY-plots}), and retains 69\% of the YSOs from \citet{Manara_TTauri}. Sources that satisfy this filter have the `flag\_pEW' quality flag in Table \ref{table:AccretionTable_extract}.
    
    \item[{C}] Combining A and B: We applied both A and B criteria to produce a very pure sample of YSOs. By doing this, we obtained an all-sky sample of 1\,945 sources (Fig. \ref{fig:SKY-plots}), and we retained 60\% of the YSOs from \citet{Manara_TTauri}. Sources that satisfy this filter have the `flag\_combined' quality flag in Table \ref{table:AccretionTable_extract}. 
\end{enumerate}

\begin{figure}
    \centering
    \includegraphics[width=\linewidth]{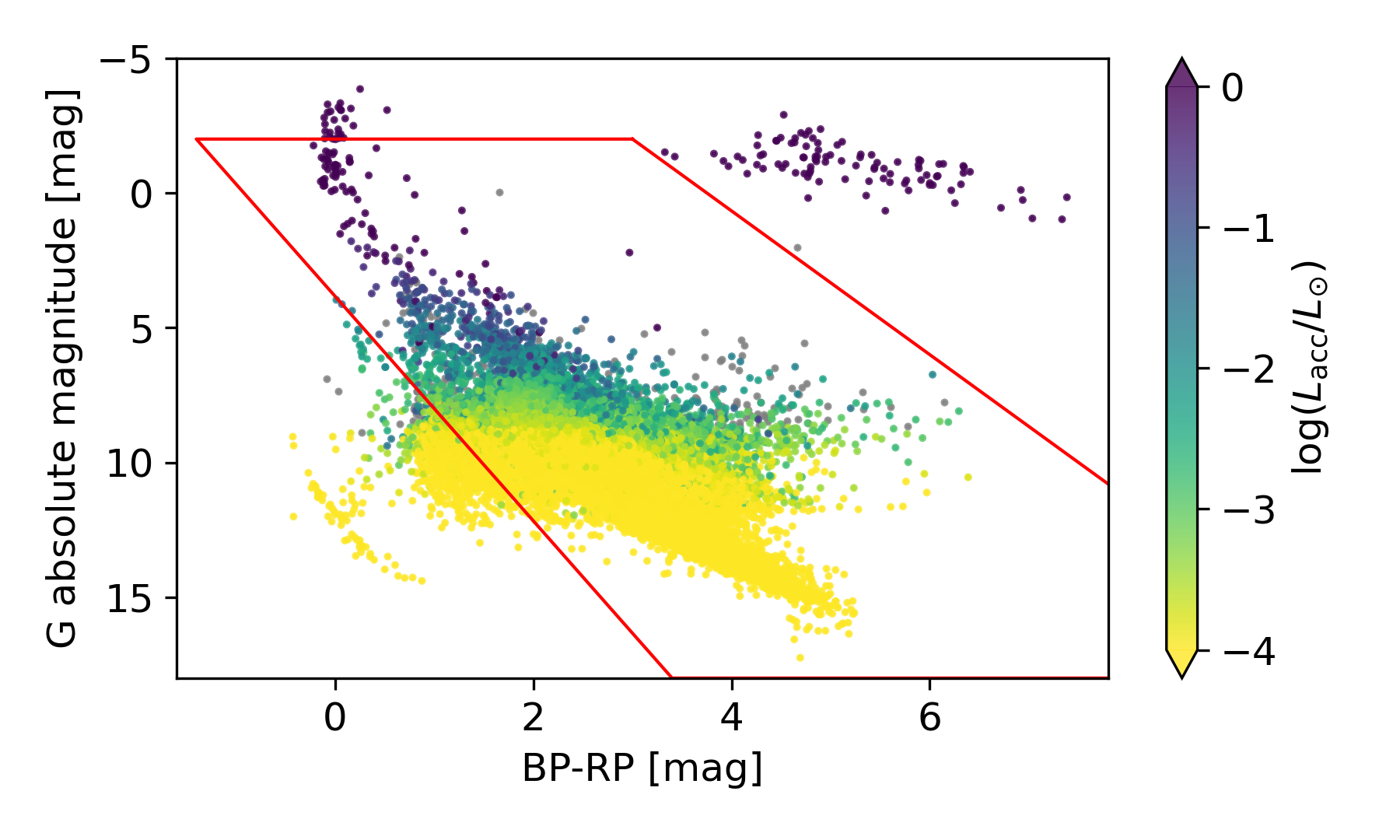}
    \caption{Colour-magnitude diagram (CMD) of all H\textalpha{} emitter candidates within 500\,pc (Table \ref{table:AccretionTable_extract}). The 1\,494 sources outside of the red lines are flagged in Sect. \ref{Sect:YSO_flags_criteria} as being incompatible with a YSO location in the CMD. The equations to reproduce the lines are reported in Appendix \ref{Appendix C}. The sources are colour-coded by accretion luminosity, using the accretion luminosity calculated using med-GSP-Phot extinction. Grey sources belong to the 1\% of the sample which does not have med-GSP-Phot extinction.}
    \label{fig:HR_cut}
\end{figure}

\begin{figure}
    \centering
    \includegraphics[width=\linewidth]{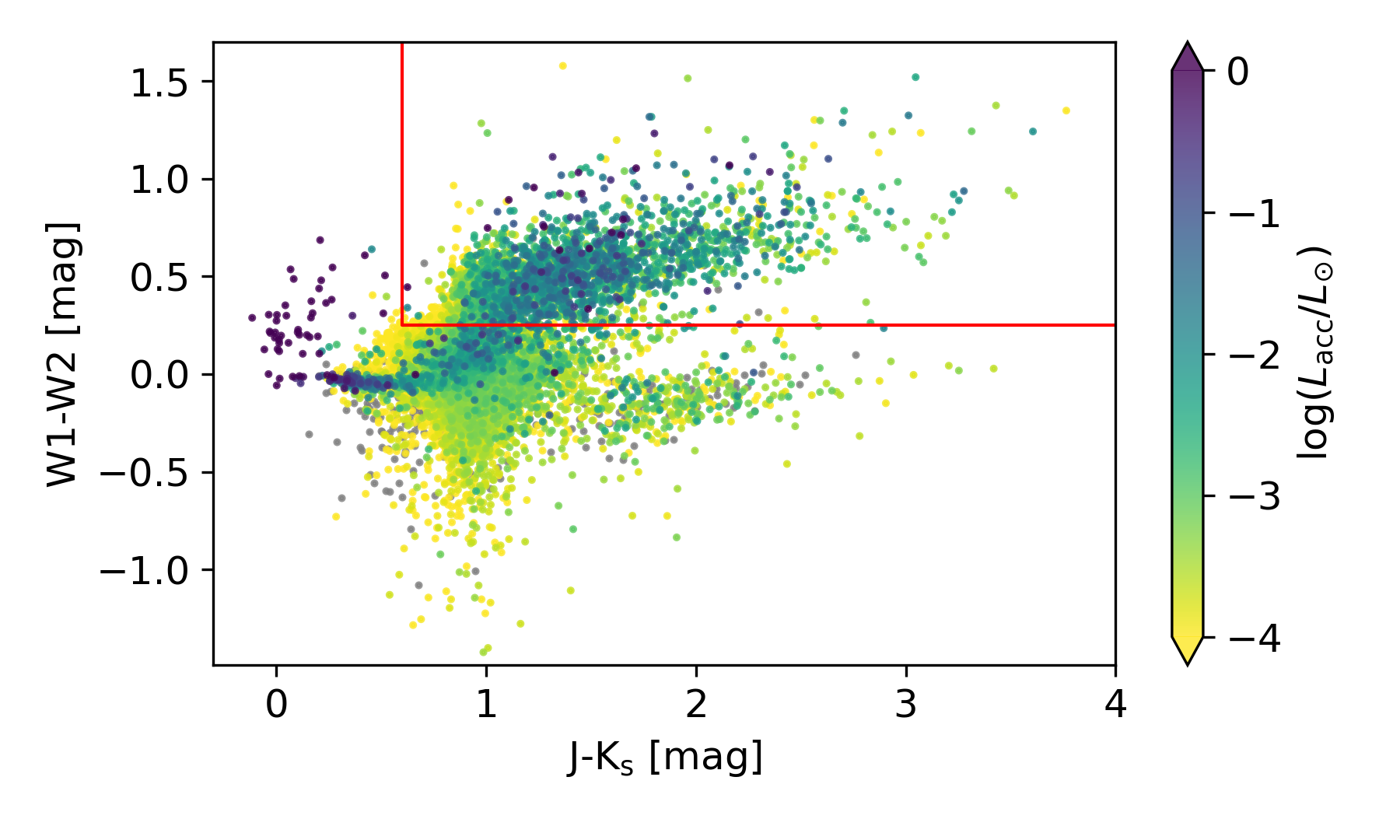}
    \caption{Colour-colour plot using 2MASS and WISE magnitudes. The sources in the plot were pre-selected using the colour-magnitude diagram of Fig. \ref{fig:HR_cut}, and by excluding upper limits.  The region to the top right ($\text{J}-\text{K\textsubscript{s}} > 0.60$ mag and $\text{W1}-\text{W2} > 0.25$ mag) is the region considered when applying selection criterion A (Sect. \ref{Sect:YSO_flags_criteria}). The sources are colour-coded by accretion luminosity, using the accretion luminosity calculated using med-GSP-Phot extinction. Grey sources belong to the 1\% of the sample which does not have med-GSP-Phot extinction.}
    \label{fig:col_col_cut}
\end{figure}

\begin{figure}
    \centering
    \includegraphics[width=\linewidth]{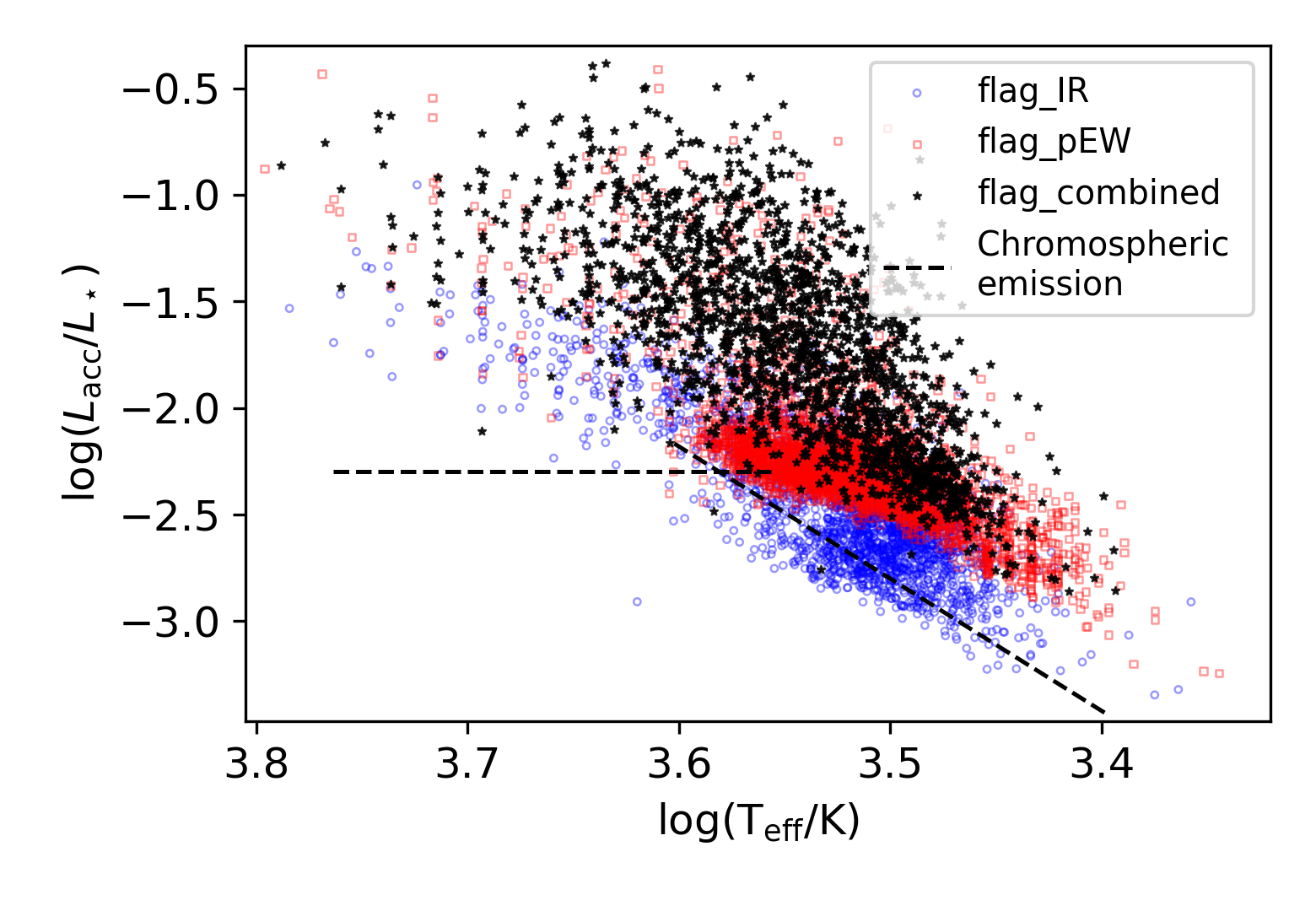}
    \caption{Accretion luminosity divided by stellar luminosity as a function of effective temperature for the all-sky YSOs with accretion luminosity and effective temperature calculated from med-GSP-Phot extinction, from samples A (flag\_IR, 4\,086 sources), B (flag\_pEW, 5\,934 sources), and C (flag\_combined, 1\,871 sources). All sources from sample C are, by definition, also included in sample A and B.
    Dashed lines trace the locus of chromospheric emission lines when they are erroneously converted into accretion luminosities, as derived in \citet{Manara_chromospheric_emission}. 8.44\% of sample A and 0.15\% of sample B are below the chromospheric emission lines.}
    \label{fig:chromospheric_emission}
\end{figure}

We note that, while the flag on IR excess (criterion A, Fig. \ref{fig:col_col_cut}) should be used when aiming for a low-contamination sample of sources with a significant amount of emission in the near IR, it does not imply that all sources where this flag is false do not have IR excess. Two main factors may cause sources with IR excess to have `flag\_IR' set to false. Firstly, the flag is false if the photometry in any one of J, $\text{K\textsubscript{s}}$, W1 or W2 is an upper limit. Secondly, IR excess could begin at longer wavelengths. For example, if we consider the 6\,170 sources of sample B (sources with strong H\textalpha{} signatures, `flag\_pEW'), only 32\% have `flag\_IR'. However, for 13\% of sample B the flag is false due to upper limits. Additionally, from the remaining 3\,435 sources we take the 278 that have non-upper limit W3 and W4 magnitudes, and search for excess at longer wavelengths. We find that 90\% have $\text{W3}-\text{W4}$\,$> 1$ (and 81\% have $\text{W3}-\text{W4}$\,$> 2$). These numbers indicate that there is a large fraction of YSOs where the IR excess only starts at longer wavelengths, but we do not apply a selection at longer wavelengths due the quality of W3 and W4 photometry.

An additional quality flag is available in Table \ref{table:AccretionTable_extract}, labelled as `flag\_above\_chromospheric\_level'. This flag indicates whether accretion is observed above the level of chromospheric emission, using the results from \citet{Manara_chromospheric_emission}. If accretion is below this level, the measured accretion luminosity and mass accretion rate could be spurious detections incorrectly derived from chromospheric emission. We derived this flag for the whole sample of sources with med-GSP-Phot stellar parameters (96\% of the total sample of Table \ref{table:AccretionTable_extract}), and found that 55\% of sources are below the chromospheric emission level. We then checked how many sources in each of samples A, B, and C are below this level (Fig. \ref{fig:chromospheric_emission}). For sample A, 8.44\% of the sample is located below the locus of chromospheric emission lines, indicating some degree of contamination. For sample B, only 0.15\% is located below the locus, and for sample C, no sources in the sample (save two) are below the locus. This is a strong indication that all sources in sample C are indeed true YSO accretors.

\begin{figure}
    \centering
    \includegraphics[width=0.9\linewidth, clip]{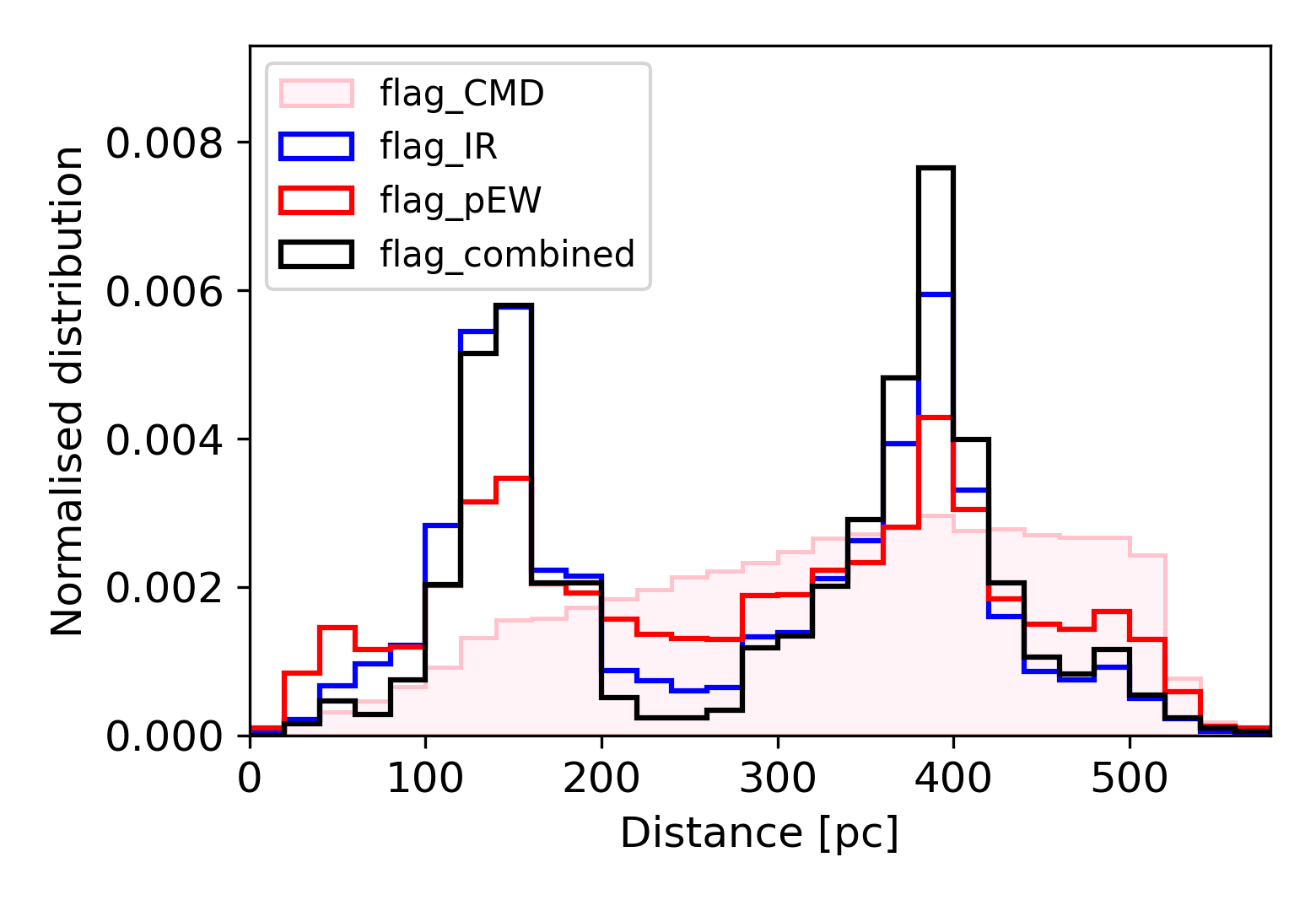}
    \caption{Histogram of distances of YSO H\textalpha{} emitter candidates including the full table with `flag\_CMD' and samples A, B, and C (these being refined subsets of YSO candidates, Sect. \ref{Sect:YSO_flags_criteria}). We use the geometric distances of \cite{geom_dist}.}
    \label{fig:distance_histogram}
\end{figure}

\begin{figure*}[htbp]
    \centering
    \includegraphics[width=\linewidth]{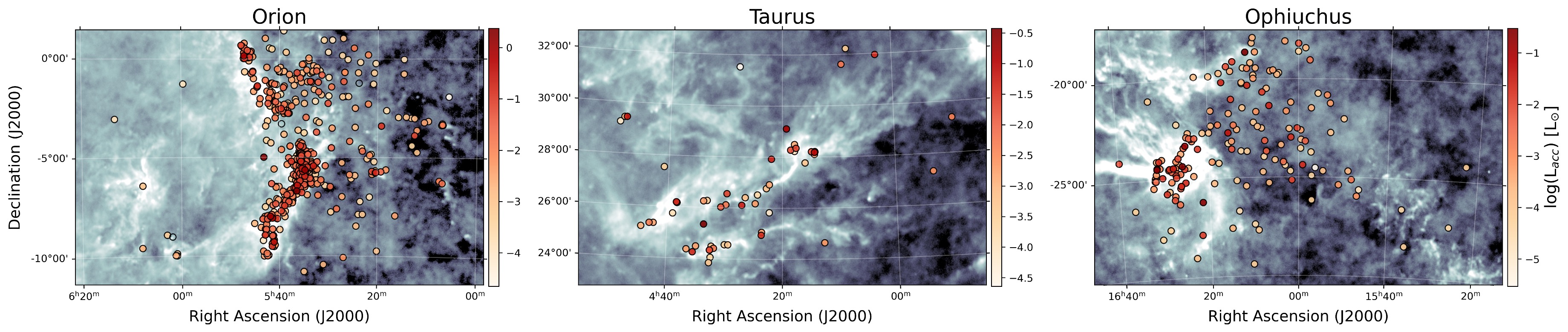}
    \caption{Detail of some star-forming regions from Fig. \ref{fig:SKY-plots} bottom panel (sample C), plotted over the Planck map at 217\,GHz \citep{Planck_collaboration}. YSOs with higher accretion luminosities can be appreciated concentrated in the more clustered areas, following the structure of the regions.}
    \label{fig:SFR_zoomed}
\end{figure*}

Fig. \ref{fig:SKY-plots} shows the three resulting samples from filters A, B, and C plotted in galactic co-ordinates, and colour-coded by their accretion luminosity. In all three cases, the regions of stronger accretion correspond to known star-forming regions, but there is a population of low-accreting disperse YSOs. Zooming into star-forming regions and plotting them on the Planck 217 GHz map \citep{Planck_collaboration} shows the spatial distribution of YSOs following the structure of dust (Fig. \ref{fig:SFR_zoomed}). While not shown here, a similar structure is observed when colour-coding by mass accretion rate. Fig. \ref{fig:distance_histogram} shows a histogram of the distances of YSO candidates including the full table with `flag\_CMD' and samples A, B, and C. For A, B, and C the distribution shows two peaks, the first corresponding mostly to the Sco-Cen and Taurus regions, and the second to Orion. The table with `flag\_CMD' shows no sign of structure as a function of distance. This illustrates the importance of applying at least one of the quality cuts described in this section for retrieving pure YSO populations.

\begin{figure*}
\includegraphics[width=\linewidth, clip]{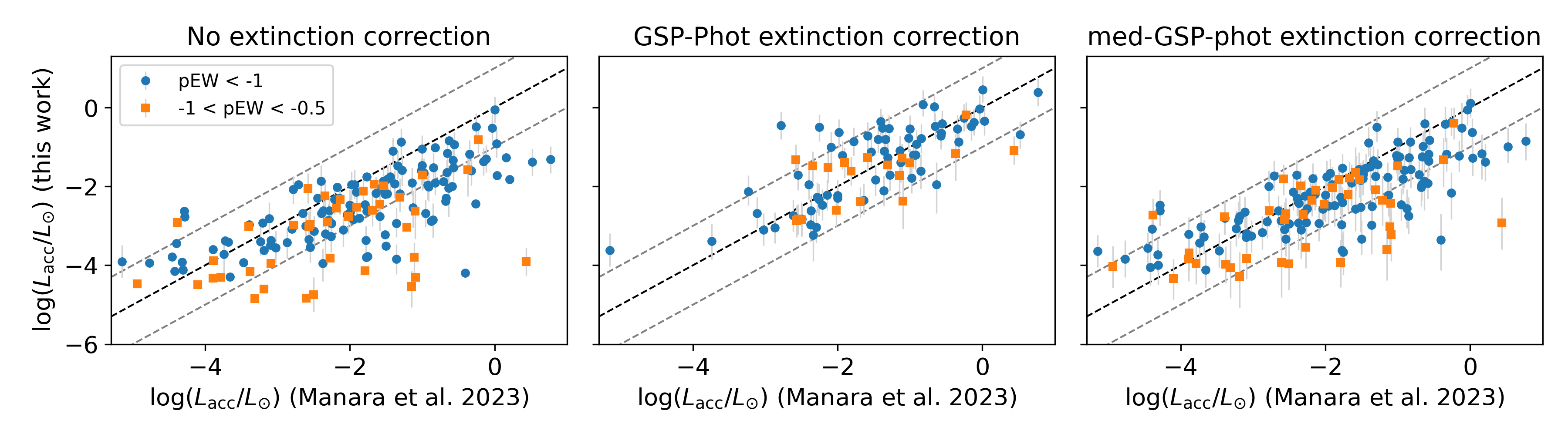}
\centering
\includegraphics[width=0.68\linewidth, clip]{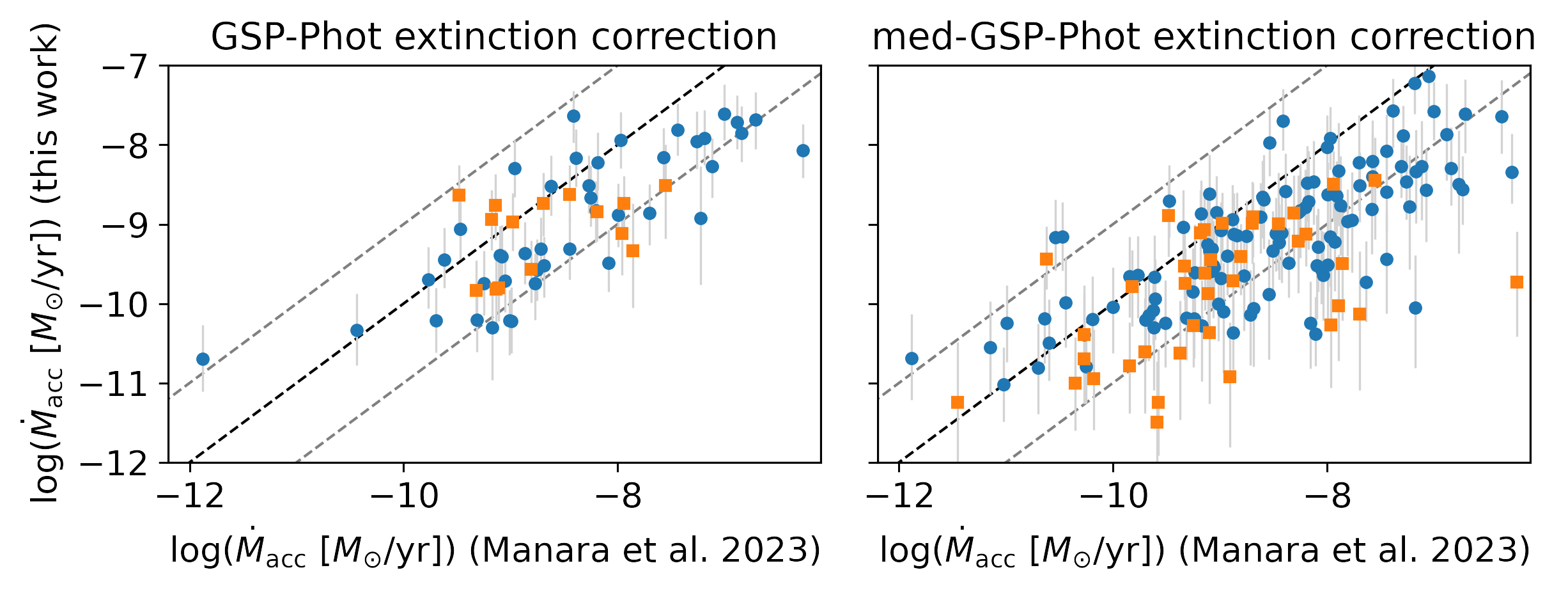}
\caption{Comparison of the accretion luminosities and mass accretion rates of this work with the values reported in \citet{Manara_TTauri} from higher-resolution spectroscopy. The dashed black lines are the 1:1 lines, and the dashed grey lines show $\pm1$ dex. \textit{Top panels}: Three plots showing accretion luminosities calculated without accounting for extinction (left), using \textit{Gaia} GSP-Phot extinction (central), and using med-GSP-Phot extinction (right). \textit{Bottom panels}: Two plots showing mass accretion rates calculated using \textit{Gaia} GSP-Phot extinction (left), and using med-GSP-Phot extinction (right).}
\label{Fig:Macc_Manara_copmarison}
\end{figure*}
\subsection{Comparison with literature values of accretion luminosity and mass accretion rate}\label{sec: lit}

We compared the accretion luminosities and mass accretion rates derived in Sect. \ref{sec:Lacc_derivation} with the accretion properties of the sample of known classical T Tauri stars from \citet[and references therein]{Manara_TTauri}. We have 341 YSOs in common, and the sample contains objects from various regions within 300\,pc, including Lupus, Taurus, Ophiuchus, Chamaeleon I, Chamaeleon II, Corona Australis, and Upper Scorpius. The stellar and accretion parameters provided by \citet{Manara_TTauri} were mainly obtained from surveys carried out with the VLT/X-Shooter instrument ($R\sim10\,000$). Fig. \ref{Fig:Macc_Manara_copmarison} shows the comparison of our accretion luminosities and mass accretion rates with these literature values from higher-resolution spectroscopy, for the different extinction corrections considered in this work (Sect. \ref{ext_section}).

Fig. \ref{Fig:Macc_Manara_copmarison} shows that the accretion luminosities and mass accretion rates estimated from the \textit{Gaia} XP spectra are accurate to within an order of magnitude, when extinction is taken into account. A better match is found for the GSP-Phot extinction correction than for the med-GSP-Phot extinction correction. This is expected, as circumstellar extinction is an important component of the extinction in YSOs and med-GSP-Phot, by definition, only traces interstellar extinction (and hence systematically supposes a lower limit to the real extinction, Sect. \ref{ext_section}). Because of this, med-GSP-Phot based accretion luminosities and mass accretion rates tend to underestimate the values obtained from higher-resolution spectroscopy. We note, however, that med-GSP-Phot extinctions are available for 99\% of the original sample of H\textalpha{} emitters within 500\,pc, as opposed to the 68\% in the case of GSP-Phot extinctions (Sect. \ref{ext_section}). This is why med-GSP-Phot based accretion luminosities were used in Figs. \ref{fig:HR_cut}, \ref{fig:col_col_cut}, \ref{fig:chromospheric_emission}, \ref{fig:SKY-plots}, and \ref{fig:SFR_zoomed}. pEW\,$<-1.0$\,nm sources are indicated in Fig. \ref{Fig:Macc_Manara_copmarison} to illustrate the behaviour of the accretion parameters derived from XP spectra when using criterion B for selecting a purer sample of YSO H\textalpha{} emitters (Sect. \ref{Sect:YSO_flags_criteria}). 

The lower limit to the mass accretion rate that we are sensitive too is $\dot{M}_\text{acc} \sim 10^{-12}$ M$_{\odot}$/yr, or log($L_\text{acc}/L_\odot) \sim -5.0$ in the case of accretion luminosity. These lower detection limits are inferred because in the comparison with literature values the latter are all above these limits (Fig. \ref{Fig:Macc_Manara_copmarison}), and the distribution of accretion properties for the whole set of H\textalpha{} emitters drops abruptly at these limits (see Sect. \ref{Sec:Accretion_all_sky}).

\begin{figure*}[h]
    \centering
    \includegraphics[width=1\linewidth]{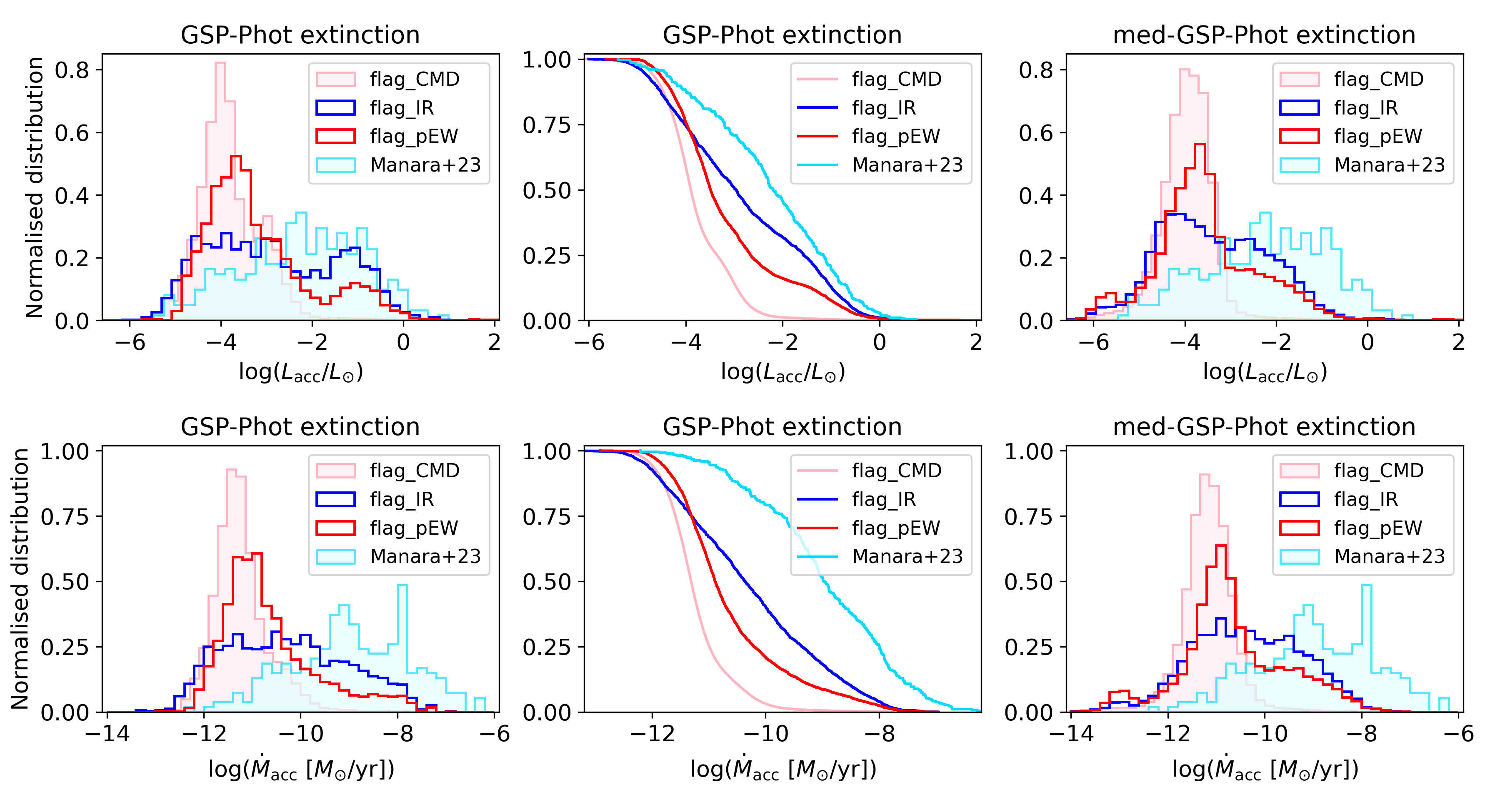}
    \caption{Distribution of accretion luminosities and mass accretion rates derived for YSO H\textalpha{} emitters within 500\,pc. Different lines trace the quality filters described in Sect. \ref{Sect:YSO_flags_criteria} (see Fig. \ref{fig:selection_diagram}). The left panels present accretion properties obtained using the GSP-Phot extinction correction and the right panels the ones obtained using the med-GSP-Phot extinction correction (Sect. \ref{ext_section}). We have added to this plot the accretion luminosities and mass accretion rates presented in \citet{Manara_TTauri} for well-characterised T Tauri stars from higher-resolution spectroscopy. The central panels are cumulative distributions of the accretion properties shown on left panels, using GSP-Phot extinction.}
    \label{fig:hist_mosaic}
\end{figure*}

\section{All-sky view of YSO accretion} \label{sec:all-sky}

In this section, we present an all-sky homogeneous view of YSO accretion within 500\,pc, using the accretion properties we obtained from \textit{Gaia} XP spectra (Table \ref{table:AccretionTable_extract}). We analyse the sky distribution of YSO accretion and correlate accretion properties to stellar parameters.

\subsection{Distribution of accretion luminosities and mass accretion rates}\label{Sec:Accretion_hists}

In Fig. \ref{fig:hist_mosaic} we show the distribution of accretion luminosities and mass accretion rates derived in Sect. \ref{sec:Lacc_derivation} for YSO H\textalpha{} emitters within 500\,pc with the quality filters of Sect. \ref{Sect:YSO_flags_criteria} (see Fig. \ref{fig:selection_diagram}). For comparison purposes we have added to this plot the accretion luminosities and mass accretion rates presented in \citet{Manara_TTauri} for well-characterised T Tauri stars from higher-resolution spectroscopy.

Fig. \ref{fig:hist_mosaic} shows that the literature sample of T Tauri stars with well-described accretion properties (i.e. \citealp{Manara_TTauri} sample) consists of stronger accretors than our YSO samples. While it is likely that the `flag\_CMD' peak at log($L_\text{acc}/L_\odot) \sim -4$ in Fig. \ref{fig:hist_mosaic} is affected by the presence of non-YSO contaminants (Sect. \ref{Sect:YSO_flags_criteria}), the `flag\_IR' and `flag\_pEW' selections also peak at this low-accretion regime, suggesting a population of YSO low accretors largely untraced by previous spectroscopic surveys.  Moreover, previous spectroscopic surveys are well contained by our IR excess selection (`flag\_IR'), indicating a bias of previous studies towards YSO accretors with IR-bright protoplanetary discs. 

We note that the YSO populations of H\textalpha{} emitters identified in this work also have different distributions depending on the filtering criteria we apply. The cumulative distributions of Fig. \ref{fig:hist_mosaic} indicate that demanding a certain level of IR excess tends to favour stronger accretors more efficiently than demanding a certain level of pEW. This is particularly evident for the strongest accretors with log($L_\text{acc}/L_\odot) > -2$. We believe this to be a consequence of two effects: more massive discs are correlated with stronger accretion (\citealp{2016A&A...591L...3M}), and discs disperse with time while accretion also diminishes with time (\citealp{Fedele10_accretion_timescale,2015A&A...576A..52R}, see also Sect. \ref{Sect:ScoCen_section}). The previously described effects are even more notorious when considering the mass accretions rates (albeit these are more uncertain in this work, as they have a higher dependence on extinction and stellar parameters). 

As is described in Sect. \ref{sec: lit}, the med-GSP-Phot based accretion properties are systematically underestimated. This can also be appreciated in Fig. \ref{fig:hist_mosaic}. Nevertheless, similar trends as observed for GSP-Phot based accretion properties can be observed for the larger sample of objects with accretion properties derived using med-GSP-Phot. The two peaked profile distribution of GSP-Phot based accretion luminosities is a direct consequence of the two peak extinction profile reported by \textit{Gaia} GSP-Phot (\citealp{extinction_GSPPHOT}) for the same sample of sources. This is an artifact of the GSP module, which tends to overestimate extinctions for sources with notable extinction\footnote{See Chapter 11.3.3 of the \textit{Gaia} DR3 Documentation, available at \url{https://gea.esac.esa.int/archive/documentation/GDR3/}}.The two-peak distributions indeed transform into smooth distributions when averaging the extinction of several sources (using med-GSP-Phot). Because of this we did not derive stellar parameters (and hence mass accretion rates) for sources that fall outside of the \citet{BHAC15} tracks in the dereddened CMD (Sect. \ref{sec:derivation_stellar_parameters}). This explains the absence of a double-peaked distribution in the mass accretion rates calculated with GSP-Phot, as sources with slightly overestimated extinctions fall outside of the \citet{BHAC15} tracks.

The sky distribution of the YSO low accretors can be observed in Fig. \ref{fig:SKY-plots}, where it is noticeable that the population of low accretors is mostly dispersed, away from star-forming regions or clustered environments of star formation. Even within star-forming regions, the YSO low accretors appear disperse and separated from the more clustered areas, where the accretion luminosities are higher (Fig. \ref{fig:SFR_zoomed}). This spatial correlation of accretion, with higher accretors appearing more clustered and structured, can be related to ongoing infall from the interstellar medium onto the protoplanetary discs (\citealp{2023EPJP..138..272K,2024A&A...683A.133G,2024A&A...691A.169W}), star-disc encounters in denser areas (\citealp{2024A&A...691A..43W}), evolution of YSOs with time towards isolation (\citealp{2023AJ....166..183V}), or ejected YSOs because of multiple system disruptions (\citealp{2024ApJ...975..207C}). It is also worth noting that there is a small known population of very old ($>20$\,Myr) still accreting M-type YSOs (know as `Peter Pan' discs, \citealp{2020ApJ...890..106S}) whose protoplanetary discs are likely primordial (\citealp{2024arXiv241205535L}). \citet{2020MNRAS.496L.111C} predicted these sources should form in the periphery of low-mass star-forming regions or in more isolated locations. It could then be that a fraction of the population of field YSO low accretors we detect are old YSOs with surviving primordial discs that move from their little-crowded forming places.

\subsection{All-sky view of accretion properties versus stellar parameters} \label{Sec:Accretion_all_sky}

Several relations have been found between the accretion properties of YSOs, their stellar parameters, and the properties of their protoplanetary discs, which have been used to describe the evolution of YSOs, their accretion mechanisms, and the transport of angular momentum in protoplanetary discs (see \citealp{Manara_TTauri} and references therein). In particular, stellar luminosities are known to positively correlate with accretion luminosities, and stellar masses to positively correlate with mass accretion rates. In this section we revisit these correlations with the largest sample ever considered of YSOs, homogeneously obtained for the whole sky.

\begin{figure*}
    \centering
    \includegraphics[width=0.99\columnwidth]{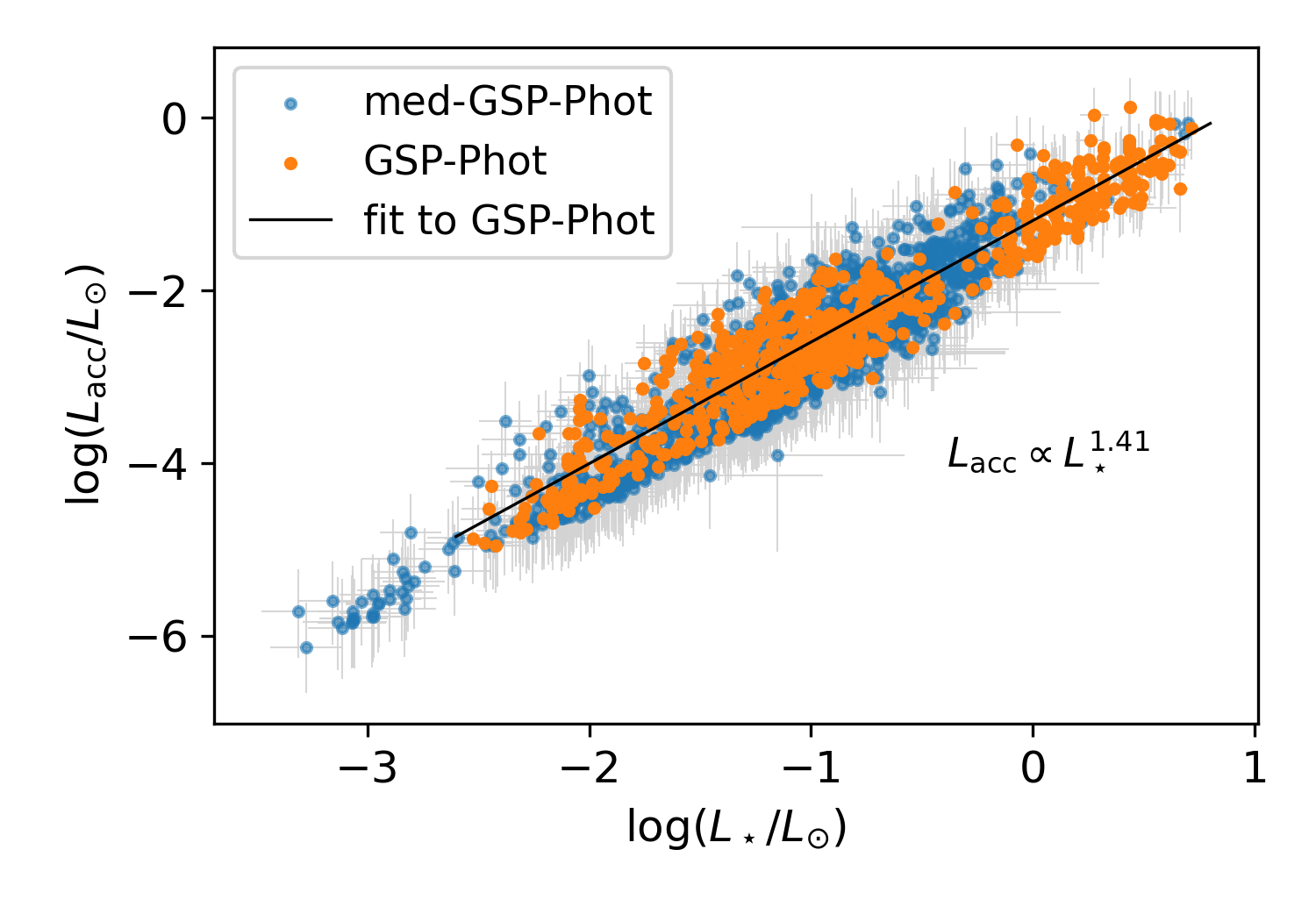}
    \includegraphics[width=0.99\columnwidth]{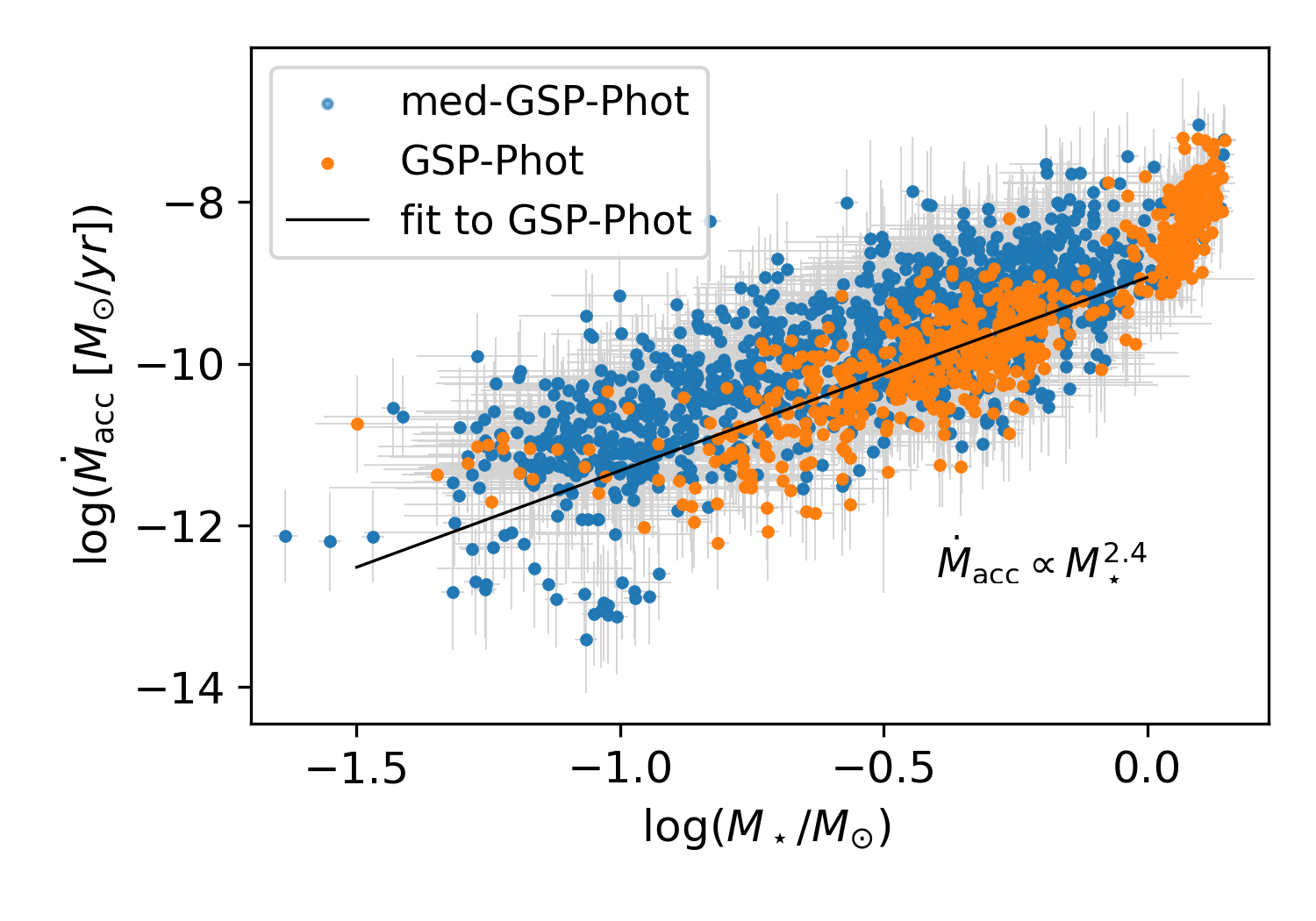}
    \caption{\textit{Left:} Accretion luminosity vs stellar luminosity for 1\,871 YSOs all-sky within 500\,pc (sample C, Sect. \ref{Sect:YSO_flags_criteria}). The 655 sources of the sample that have GSP-Phot extinction are plotted according to their values of accretion luminosity and stellar luminosity derived using this extinction. The black line is the fit for the GSP-Phot values of these 656 sources. The value of the slope in the log-log plane is $1.41 \pm 0.02$. The remaining sources of sample C that do not have GSP-Phot extinction are plotted using the values calculated using med-GSP-Phot. The results of fitting to values calculated with med-GSP-Phot are reported in Table \ref{tab:Lacc_Lstar_fit}. \textit{Right:} Mass accretion rate vs stellar mass for the same sample of 1\,871 YSOs. The 656 sources of the sample that have GSP-Phot extinction are plotted according to their values derived using this extinction. The black line is the fit to the GSP-Phot values, for sources with stellar mass $M_\star < 1.0$ M$_\odot$, and with $\dot{M}_\text{acc}>10^{-12}$ M$_\odot/$yr. The value of the slope in the log-log plane is $2.4 \pm 0.1$. Again, the sources of sample C that do not have GSP-Phot extinction are plotted using the values calculated using med-GSP-Phot.}
    \label{fig:Lstar_Lacc_Macc_relation}
\end{figure*}

In order to minimise the number of contaminants, we took sample C (`flag\_combined', Sect. \ref{Sect:YSO_flags_criteria}), which consists of 1\,945 sources. From these, we considered the 656 sources for which we have derived stellar parameters, accretion luminosities, and mass accretion rates from GSP-Phot extinction. We then looked at the correlation of accretion luminosity with stellar luminosity and mass accretion rate with stellar mass, shown in Fig. \ref{fig:Lstar_Lacc_Macc_relation}. Fitting the logarithmic relation between accretion luminosity and stellar luminosity (using the Bayesian method from \citet{linmix} to account for uncertainties in both quantities), we obtained $L_{\text{acc}}\propto L_\star ^{1.41\pm 0.02}$. We then looked for a correlation between mass accretion rate and stellar mass, shown in Fig. \ref{fig:Lstar_Lacc_Macc_relation}. As we have no empirical calibration for sources with $\dot{M}_\text{acc}<10^{-12}$ M$_\odot/$yr, we exclude the sources in this limit. We set a conservative upper limit of 1.0\,M$_\odot$ for the fit to avoid entering the massive end regime of the \citet{BHAC15} tracks (Sect. \ref{sec:derivation_stellar_parameters}). We fit a logarithmic relation, again accounting for uncertainties, and obtained that $\dot{M}_\text{acc} \propto M_{\star}^{2.4 \pm 0.1}$. We repeated the procedure for the samples A and B of YSOs (respectively, `flag\_IR' and `flag\_pEW'), and again repeated it using the values calculated using med-GSP-Phot extinction (we note that for med-GSP-Phot extinction we did not impose any constraints on the stellar mass). Fig. \ref{fig:Lstar_Lacc_Macc_relation} shows med-GSP-Phot values in blue for the sources of sample C that do not have GSP-Phot extinction. The results of the fit for the different samples and extinction regimes are tabulated in Table \ref{tab:Lacc_Lstar_fit}. Comparing the values of the slopes in the logarithmic plane calculated with the different methods, we find a standard deviation of 0.05 for the $L_{\text{acc}} - L_\star$ relation and of 0.13 for the $\dot{M}_\text{acc}$ - $M_\star$ relation. The different methods therefore give consistent results, especially for the $L_{\text{acc}} - L_\star$ relation.

\begin{table*}[h]
    \caption{Results of the power law fits $\log(L_{\text{acc}}) = m\cdot\log(L_\star) + c$ and $\log(\dot{M}_\text{acc}) = m\cdot\log(M_\star) + c$, for the different samples of YSOs defined in Sect. \ref{Sect:YSO_flags_criteria}, and for values calculated from both GSP-Phot extinction and med-GSP-Phot extinction.}
    \centering
    \small
    \begin{tabular}{c|c|c|c c|c c}
    \hline\hline\\[-0.9em]   
    Flag & Extinction & N. sources & $m_{L_\text{acc}}$ & $c_{L_\text{acc}}$ & $m_{\dot M_\text{acc}}$ & $c_{\dot M_\text{acc}}$ \\   
    \hline  \noalign{\smallskip}                     
    IR & GSP-Phot & 1564 & $1.51 \pm 0.01$ & $-1.42 \pm 0.02$ & $2.5 \pm 0.1$ & $-9.31 \pm 0.05$\\
    pEW & GSP-Phot & 2836 & $1.53 \pm 0.01$ & $-1.29 \pm 0.01$ & $2.25 \pm 0.06$ & $-9.72 \pm 0.03$\\
    combined & GSP-Phot & 656 & $1.41 \pm 0.02$ & $-1.19 \pm 0.02$ & $2.4 \pm 0.1$ & $-8.93 \pm 0.05$\\
    \hline \noalign{\smallskip}
    IR & med-GSP-Phot & 4086 & $1.57 \pm 0.01$ & $-1.37 \pm 0.01$ & $2.38 \pm 0.04$& $-8.87 \pm 0.02$\\
    pEW & med-GSP-Phot & 5934 & $1.53 \pm 0.01$ & $-1.30 \pm 0.01$ & $2.20 \pm 0.04$ & $-9.29 \pm 0.02$ \\
    combined & med-GSP-Phot & 1871 & $1.50 \pm 0.02$ & $-1.16 \pm 0.02$ & $2.19 \pm 0.05$ & $-8.66 \pm 0.03$ \\
    \hline
    \end{tabular}
    \label{tab:Lacc_Lstar_fit}
\end{table*}

We now compare the parameters of Table \ref{tab:Lacc_Lstar_fit} with those found in previous studies. In this comparison, we note that we are tracing a different population of YSOs (Fig. \ref{fig:hist_mosaic}), and that different stellar evolutionary models might have been used to obtain stellar parameters (this does not affect $L_\text{acc}$). For the $L_\text{acc}$ - $L_\star$ correlation in the logarithmic plane, our values are in the range 1.41 to 1.57 for the slope and between $-1.42$ and $-1.16$ for the intercept, with best values (those of sample C with GSP-Phot extinction) of $1.41 \pm 0.02$ and $-1.19 \pm 0.02$, respectively. \citet{Almendros-Abad24} fit the power law correlation for sources in four different star-forming regions. For the $L_\text{acc}$ - $L_\star$ relationship, they find values of the slope in the logarithmic plane in the range 1.19 to 1.76, and intercept in the range $-1.65$ to $-1.1$. Our values all fall within this range. \citet{Manara2017} find a slope of $1.9\pm0.1$, and \citet{Tilling2018} present a theoretical study which uses simplified stellar evolution calculations for stars subject to a time-dependent accretion history, obtaining a slope of 1.7. Compared to the latter two studies, we find a flatter power law. For the $\dot{M}_\text{acc}$ - $M_\star$ correlation, we obtain values of the slope in the range $2.19$ to $2.5$, and of the intercept in the range $-9.72$ to $-8.66$, with best (sample C) values of $2.4\pm0.1$ and $-8.93 \pm 0.05$. Our power law is within error of the one found in \citet{Manara2017}, which has a slope of $2.3\pm0.3$, but slightly steeper than that of \citet{Almendros-Abad24}, where they find slopes in the range 1.33 to 2.08 (and intercept in the range $-9.17$ to $-7.76$). In conclusion, our $L_\text{acc}$ - $L_\star$ correlation is consistent with the results of \citet{Almendros-Abad24}, while the $\dot{M}_\text{acc}$ - $M_\star$ power law is broadly consistent with \citet{Manara2017} but steeper than the one found in \citet{Almendros-Abad24}.

\section{Evolution of YSO accretion with age} \label{Sect:ScoCen_section}

\subsection{Evolution of accretion with age in Sco-Cen} \label{sec:ScoCen_ages}

\begin{figure*}
    \centering
    \includegraphics[width=0.88\linewidth]{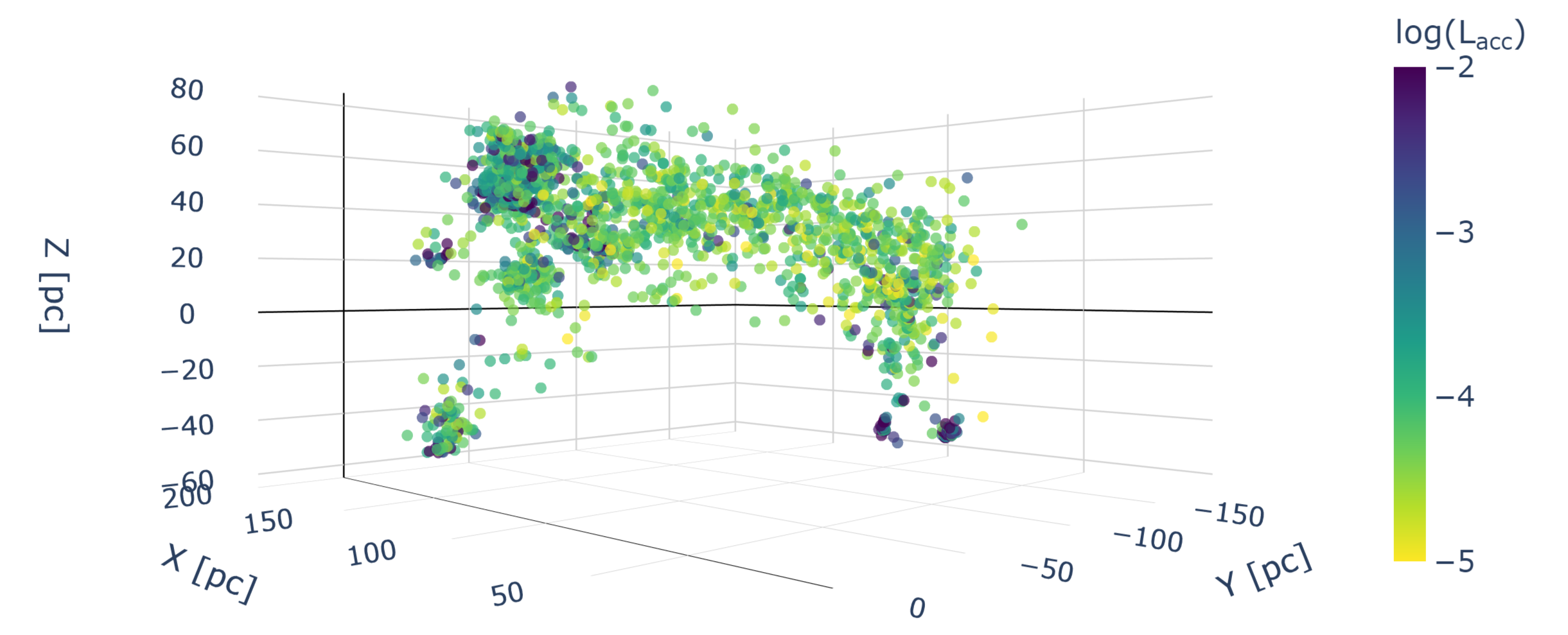}
    \caption{Three-dimensional spatial distribution of 1\,954 members of the 34 Sco-Cen clusters identified by \citet{ScoCen_SigMA_sources}, colour-coded by accretion luminosity. The co-ordinates are provided in \cite{ScoCen_SigMA_sources} and are such that the Sun is at (0,0,0) and the Z = 0 plane is parallel to the Galactic plane. For better visualisation, see the interactive 3D version \href{https://www.aanda.org/articles/aa/olm/2025/07/aa53539-24/aa53539-24.html}{online}.}
    \label{fig:ScoCen-3D}
\end{figure*}

\begin{table*}[h!]
\caption{Fraction of accreting stars, median accretion luminosity, and median mass accretion rate for 34 Sco-Cen clusters.}             
\label{table:ScoCen}      
\centering 
\resizebox{\textwidth}{!}{\begin{tabular}{c l c c c c c}
\hline\hline\\[-0.9em]   
\multicolumn{3}{c}{Sco-Cen clusters from \citet{ScoCen_SigMA_ages}}& & & \multicolumn{2}{c}{Median values per cluster}\\
    \cline{1-3}  
    \cline{6-7}\\[-0.8em] 
SigMA & Group & Age B-BpRp [Myr] & N. sources with pEW & Accretors [\%]  &  $L_{\text{acc}}$  [$L_\odot$] & $\dot M_\text{acc}$ [$M_\odot$ yr$^{-1}$]\\   
\hline  \noalign{\smallskip}                     
   
1 & $\rho$ Oph/L1688	& 3.2 (3.1,  3.4)& 406	& $15.0 \pm 1.9$		& 1.51E$-3$ (4.81E$-4$, 7.97E$-4$)  &  1.86E$-10$ (8.50E$-11$, 5.70E$-10$) \\
2 & $\nu$ Sco & 3.9 (3.4,  4.1)& 127 & $11.0 \pm 2.9$  & 6.89E$-4$ (2.02E$-4$, 8.07E$-3$)  &  6.50E$-11$ (4.25E$-11$, 4.84E$-10$) \\
3 & $\delta$ Sco & 6.4 (5.8,  7.5)& 594 & $8.6 \pm 1.2$ & 2.75E$-4$ (1.00E$-4$, 2.17E$-3$)  &  3.88E$-11$ (1.47E$-11$, 1.86E$-10$) \\
4 & $\beta$ Sco & 4.8 (4.6,  5.3)& 241 & $7.5 \pm 1.8$ & 2.43E$-4$ (1.07E$-4$, 4.46E$-4$)  &  2.93E$-11$ (1.86E$-11$, 5.86E$-11$) \\
5 & $\sigma$ Sco & 6.0 (5.8,  6.3)& 411 & $3.1 \pm 0.9$  & 1.08E$-4$ (5.40E$-5$, 1.86E$-4$)  &  1.42E$-11$ (8.88E$-12$, 4.16E$-11$) \\
6 & Antares & 6.0 (5.8,  6.5)& 429 & $5.9 \pm 1.1$  & 2.61E$-4$ (1.26E$-4$, 1.26E$-3$)  &  3.84E$-11$ (2.12E$-11$, 1.59E$-10$) \\
7 & $\rho$ Sco & 7.9 (7.4,  8.3)& 197 & $2.0 \pm 1.0$  & 1.12E$-4$ (9.90E$-5$, 1.42E$-4$)  &  2.01E$-11$ (9.39E$-12$, 3.07E$-11$) \\
8 & Scorpio-Body & 9.1 (8.8,  9.5)& 315 & $3.5 \pm 1.1$ & 5.50E$-5$ (4.30E$-5$, 9.10E$-5$)  &  1.10E$-11$ (6.20E$-12$, 1.41E$-11$) \\
9 & US-foreground & 12.5 (12.1, 13.1)& 233 & $3.9 \pm 1.3$ & 6.20E$-5$ (2.20E$-5$, 8.80E$-5$)  &  9.27E$-12$ (4.52E$-12$, 1.19E$-11$) \\
10 & V1062-Sco & 10.0 (9.8, 10.5)& 808 & $2.4 \pm 0.5$  & 9.00E$-5$ (6.00E$-5$, 1.40E$-4$)  &  1.11E$-11$ (8.92E$-12$, 2.41E$-11$) \\
11 & $\mu$ Sco & 9.6 (9.0, 10.5)& 40 & $2.5 \pm 2.5$  &  &  \\
12 & Libra-South & 12.4 (10.8, 13.3)& 61 & 0  &  &  \\
13 & Lupus 1-4 & 4.3 (4.1,  5.2)& 188 & $25.5 \pm 3.7$  & 1.99E$-3$ (4.55E$-4$, 1.21E$-2$)  &  2.32E$-10$ (5.21E$-11$, 9.12E$-10$) \\
14 & $\eta$ Lup & 9.8 (9.1, 10.1)& 657 & $3.7 \pm 0.7$ & 5.90E$-5$ (4.70E$-5$, 1.93E$-4$)  &  1.01E$-11$ (6.90E$-12$, 5.57E$-11$) \\
15 & $\phi$ Lup & 9.9 (9.4, 10.3)& 947 & $3.9 \pm 0.6$ & 8.30E$-5$ (4.90E$-5$, 2.42E$-4$)  &  1.12E$-11$ (7.97E$-12$, 2.51E$-11$) \\
17 & $e$ Lup & 11.0 (10.7 11.5)& 431 & $2.3 \pm 0.7$  & 4.40E$-5$ (3.90E$-5$, 6.70E$-5$)  &  9.10E$-12$ (5.71E$-12$, 1.16E$-11$) \\
18 & UPK606 & 8.5 (8.0,  8.9)& 110 & $6.4 \pm 2.4$  & 1.03E$-4$ (8.30E$-5$, 2.74E$-4$)  &  2.64E$-11$ (1.59E$-11$, 4.06E$-11$) \\
19 & $\rho$ Lup & 9.4 (8.9, 10.2)& 211 & $7.1 \pm 1.8$ & 3.40E$-5$ (2.20E$-5$, 5.40E$-5$)  &  6.84E$-12$ (5.08E$-12$, 1.35E$-11$) \\
20 & $\nu$ Cen & 9.5 (8.8, 12.1)& 1401 & $3.9 \pm 0.5$  & 6.40E$-5$ (3.80E$-5$, 1.00E$-4$)  &  9.23E$-12$ (6.28E$-12$, 1.92E$-11$) \\
21 & $\sigma$ Cen & 9.5 (8.8, 9.9)& 1550 & $3.5 \pm 0.5$ & 5.50E$-5$ (3.00E$-5$, 1.89E$-4$)  &  8.98E$-12$ (5.12E$-12$, 2.23E$-11$) \\
22 & Acrux & 7.3 (6.9, 7.4)& 349 & $4.6 \pm 1.1$ & 6.30E$-5$ (4.40E$-5$, 2.37E$-4$)  &  1.14E$-11$ (6.02E$-12$, 2.96E$-11$) \\
23 & Musca-foreground & 6.9 (6.7, 7.4)& 87 & $3.4 \pm 2.0$ &  &  \\
24 & $\epsilon$ Cham & 5.4 (5.0, 6.0)& 35 & $8.6 \pm 4.9$ &  &  \\
25 & $\eta$ Cham & 6.1 (5.5, 6.9)& 28 & $7.1 \pm 5.0$ &  &  \\
26 & B59 & 3.0 (2.1, 3.1)& 22 & $40.9 \pm 13.6$ & 2.91E$-2$ (6.69E$-3$, 4.81E$-2$)  &  2.52E$-09$ (4.30E$-10$, 2.60E$-09$) \\
27 & Pipe-North & 10.1 (8.4, 10.8)& 38 & $7.9 \pm 4.6$ &  &  \\
28 & $\theta$ Oph & 9.5 (8.9, 10.5)& 85 & $1.2 \pm 1.2$  &  &  \\
29 & CrA-Main & 6.1 ( 5.5, 8.5)& 77 & $11.7 \pm 3.9$ & 3.06E$-4$ (1.66E$-4$, 6.64E$-3$)  &  4.40E$-11$ (2.72E$-11$, 8.23E$-10$) \\
30 & CrA-North & 6.6 (5.8, 6.9)& 278 & $7.2 \pm 1.6$  & 1.27E$-4$ (6.50E$-5$, 2.25E$-3$)  &  2.54E$-11$ (1.13E$-11$, 1.81E$-10$) \\
31 & Scorpio-Sting & 8.5 (8.1, 9.0)& 101 & $5.9 \pm 2.4$  & 4.10E$-5$ (2.40E$-5$, 5.30E$-5$)  &  6.24E$-12$ (5.25E$-12$, 6.91E$-12$) \\
32 & Centaurus-Far & 5.9 (5.6, 7.1)& 69 & $1.4 \pm 1.4$ &  &  \\
33 & Chamaeleon-1 & 3.0 (2.6, 3.8)& 142 & $26.1 \pm 4.3$ & 9.54E$-3$ (1.73E$-3$, 2.91E$-2$)  &  1.12E$-09$ (1.16E$-10$, 3.07E$-09$) \\
34 & Chamaeleon-2 & 2.8 (2.4, 3.1)& 43 & $44.2 \pm 10.1$  & 4.70E$-3$ (1.64E$-3$, 2.75E$-2$)  &  4.55E$-10$ (2.03E$-10$, 1.57E$-09$) \\
35 & L134/L183 & 5.7 (5.1, 6.6)& 22 & 0 &  &  \\

\hline       
\end{tabular}}

\tablefoot{The Sco-Cen clusters are the ones identified in \citet{ScoCen_SigMA_sources}. Ages are the \citet{BHAC15} Bp-Rp ages calculated in \citet{ScoCen_SigMA_ages}. The median values reported here are calculated using the med-GSP-Phot extinction correction and considering all sources in each cluster that are in Table \ref{table:AccretionTable_extract} and have pEW\,$<-1.0$\,nm. The median values reported here do not correspond to the ones presented in Fig. \ref{fig:timescales_scocen}, which were corrected for stellar mass by only considering sources in the mass range 0.08-0.25 M$_\odot$. The 16\textsuperscript{th} and 84\textsuperscript{th} percentiles are reported in brackets. Clusters have no median values if they have fewer than four members that can be used to calculate the median. The full table, including the mass-corrected medians, and the value of the $\dot{M}_{\text{acc}}$-$M_{\star}$ and $L_{\text{acc}}$-$L_{\star}$ correlations for each cluster (Fig. \ref{figure_slope_change}), is available at the CDS.}

\end{table*}

In this section, we analyse how the accretion of material in YSOs, from the protoplanetary discs onto the forming stars, evolves with age. However, YSO ages are notoriously hard to characterise, often being highly model dependent and uncertain (e.g. \citealp{2010ARA&A..48..581S,2024NatAs...8..216M,2024A&A...690A..16R}). To circumvent this limitation, we considered the work of \citet{ScoCen_SigMA_ages}, who have derived ages homogeneously for 37 clusters in the Sco-Cen star formation complex (identified in \citealp{ScoCen_SigMA_sources}). Therefore, the ages presented in this section might suffer from uncertainty and model systematics in absolute terms, but relative ages are more precise and they allow us to compare the accretion properties of clusters through YSO evolution. We discarded the clusters `Norma-North', `Oph-Southeast', and `Oph-NorthFar' as these are found in \citet{ScoCen_SigMA_ages} to be unrelated to Sco-Cen. We note that the classical old regions of Upper-Centaurus-Lupus (UCL), and Lower-Centaurus-Crux (LCC) appear here subdivided into smaller groups (\citealp{ScoCen_SigMA_sources}). 

The 34 Sco-Cen stellar clusters identified in \citet{ScoCen_SigMA_sources} contain a total of 12\,972 stars. \citet{ScoCen_SigMA_ages} provide four age estimates for each cluster, determined with two different evolutionary models, PARSEC v1.2S (\citealp{parsecv1.2_1}) and that presented in \citet{BHAC15}, and using two different colour-magnitude diagrams, ($M_G$ vs
$G_{\text{BP}}-G_{\text{RP}}$) and ($M_G$ vs
$G-G_{\text{RP}}$). In this work we used the ages from \citet{BHAC15} models and $M_G$ versus
$G_{\text{BP}}-G_{\text{RP}}$, but for consistency we examined that similar results are obtained using the other three age estimates. We cross-matched the members of the 34 clusters with our table of accretion properties and stellar parameters (Table \ref{table:AccretionTable_extract}). Fig. \ref{fig:ScoCen-3D} shows the three-dimensional spatial distribution of the 1\,954 members of Sco-Cen for which we have calculated an accretion luminosity using med-GSP-Phot extinction, colour-coded by accretion luminosity. As in Sect. \ref{Sec:Accretion_hists}, in Sco-Cen a clear spatial differentiation in accretion strength can be appreciated, with some small regions concentrating the higher accretors, and with the more dispersed YSOs being mostly low accretors.

\begin{figure}[h!]
    \centering
    \includegraphics[width=1\linewidth]{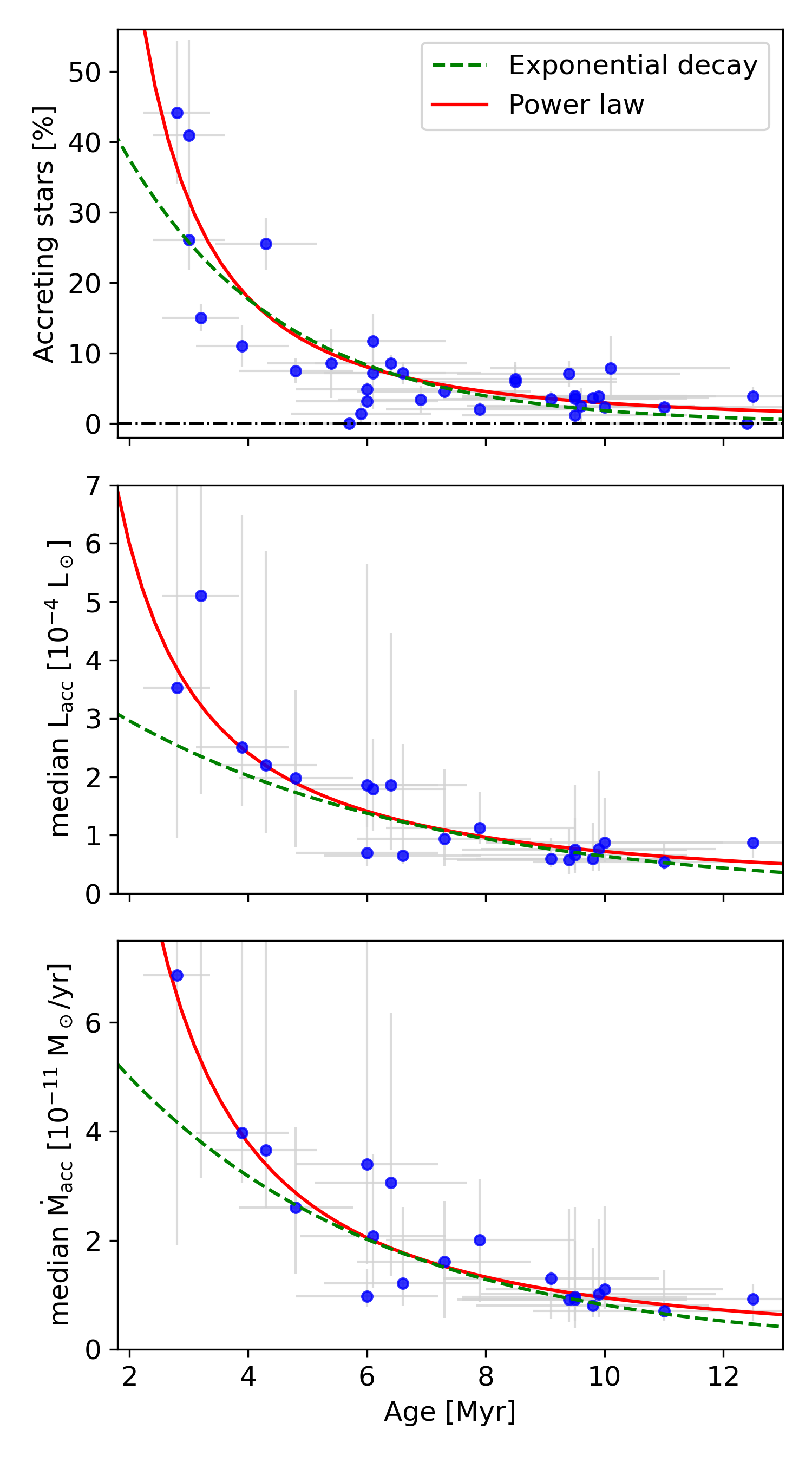}
    \caption{Evolution of accretion in Sco-Cen clusters with age. Each point represents one of the 34 Sco-Cen clusters whose ages were calculated in \citet{ScoCen_SigMA_ages}. The ages shown here are the \citet{BHAC15} Bp-Rp ages of that work. In each panel, we show two fits: an exponential decay (Eq. \ref{eqn:timescale_fraction}) and a power law (Eq. \ref{eqn:power_law}). \textit{Top panel:} we show the percentage of YSO accretors per cluster $f_{\text{acc}}$. The accretion timescale obtained from fitting the exponential is $\tau_{\text{acc}} = 2.7 \pm 0.4$\,Myr. \textit{Bottom two panels:} median accretion luminosity and the median mass accretion rate of the accretors (sources with pEW\,$<-1.0$\,nm) in each cluster. Only the cluster members with stellar mass in the range $0.08<M_\star<0.25$ M$_\odot$ were considered for the medians, to remove dependencies with stellar mass. Only clusters with four or more accretors in this mass range were considered. All fit parameters are reported in Table \ref{Table:scocen_parameters}.}
    \label{fig:timescales_scocen}
\end{figure}

\begin{figure}[h!]
    \centering
    \includegraphics[width=\linewidth]{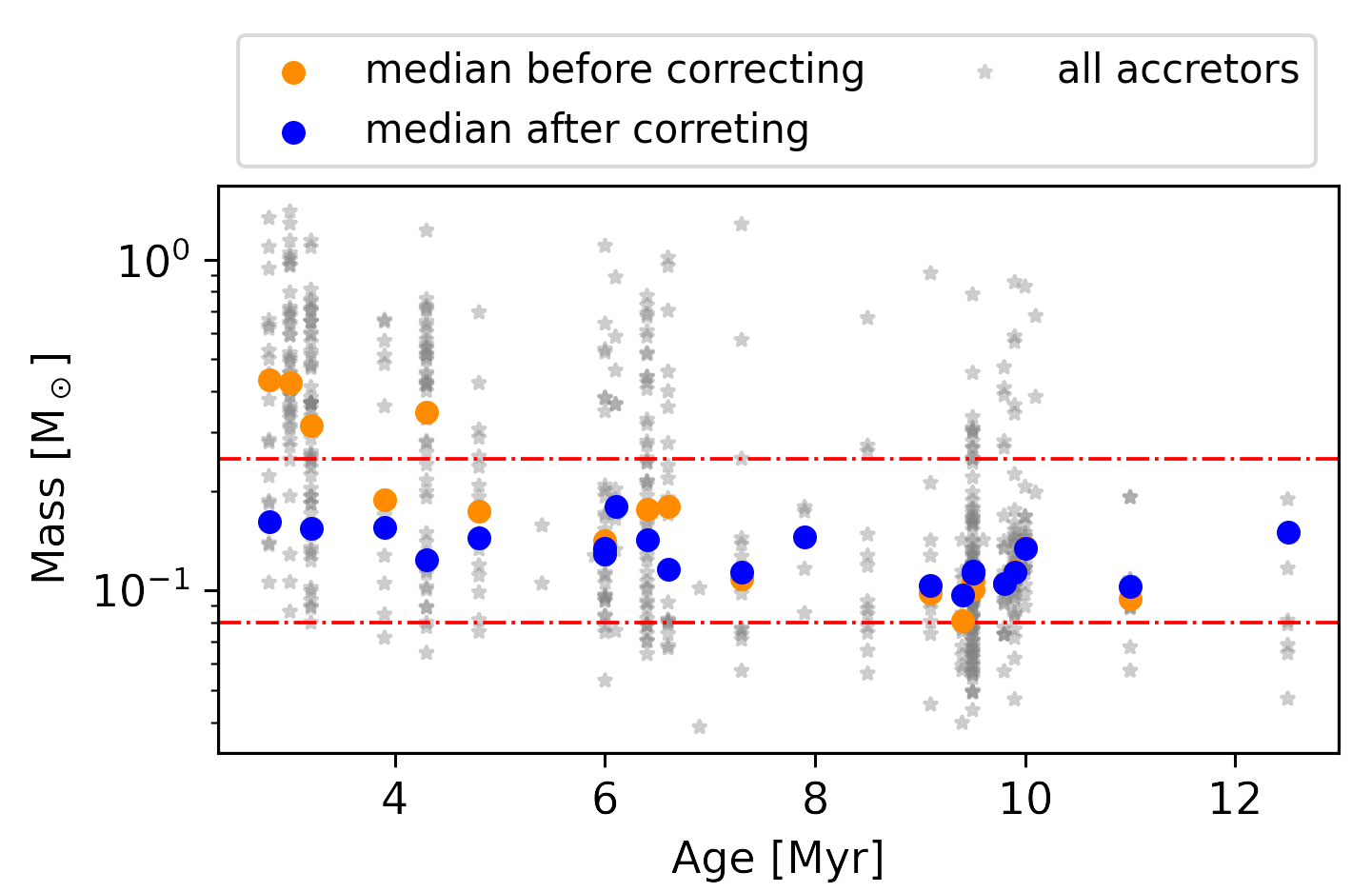}
    \caption{Stellar mass distribution of accretors in the Sco-Cen clusters vs cluster age (\citealp{BHAC15}-BpRp). The plot shows the stellar masses of all cluster members with pEW\,$<-1.0$\,nm as derived in Sect. \ref{sec:derivation_stellar_parameters} (Table \ref{table:AccretionTable_extract}), and the cluster median both before (orange) and after (blue) applying the $0.06<M_\star<0.22$ M$_\odot$ cut (red lines). The median mass is only shown if there are at least four cluster members in the chosen mass range. We note this filtering by stellar mass results in similar median stellar masses for all clusters.}
    \label{fig:Mass_distribution_scocen}
\end{figure}

For each cluster, we measured the number ($N_{\text{acc}}$) of sources that have pEW\,$<-1.0$\,nm in Table \ref{table:AccretionTable_extract}. Similarly to what we did to obtain sample B (Sect. \ref{Sect:YSO_flags_criteria}), pEW\,$<-1.0$\,nm allows us to obtain a sample of accreting YSOs that includes fewer contaminants, providing the best compromise between accuracy and completeness (Appendix \ref{Appendix B}). We then determined the total number (N) of sources in each cluster that have pEW values, and from this we calculated the fraction of sources per cluster with pEW\,$<-1.0$\,nm. We refer to this fraction as the fraction of accretors per cluster, $f_{\text{acc}}$. We estimated the uncertainty in $f_{\text{acc}}$ as $\sqrt{N_{\text{acc}}}/N$, following \citet{Fedele10_accretion_timescale}. The values of $f_{\text{acc}}$ of each cluster are presented in Table \ref{table:ScoCen}.

The top panel in Fig. \ref{fig:timescales_scocen} shows the fraction of accretors per cluster as a function of cluster age. We fit the observed decay in $f_{\text{acc}}$ with age with two different functions: an exponential decay of the form 
\begin{equation}
    f_{\text{acc}} = C\cdot \exp{(-t/\tau_{\text{acc}})},
    \label{eqn:timescale_fraction}
\end{equation}

and a power law of the form
\begin{equation}
    f_{\text{acc}} = k\cdot t^\alpha, 
    \label{eqn:power_law}
\end{equation}

where t is the age in millions of years. Both fits were performed using the method from \citet{linmix}, accounting for uncertainties in $f_{\text{acc}}$ and for a conservative 20\% error in the ages. Values of the parameters of the fits and of the reduced $\chi^2$ squared values are reported in Table \ref{Table:scocen_parameters}. Based on the $\chi_\text{red}^2$ values as well as visual inspection of the fits, it appears that a power law is a better fit for the evolution of the fraction of accretors, although the value of the parameter $k$ means that the fit might not be valid at ages lower than the range we consider. For comparison with literature, we report the value of the accretion timescale obtained from the exponential decay, which is $\tau_{\text{acc}} = (2.7 \pm 0.4)$\,Myr. This result is within error of the mass accretion timescale of 2.3 Myr calculated in \citet{Fedele10_accretion_timescale} using a similar approach on a smaller dataset. From the results of our power law fit, we calculate that the percentage of accretors is 70\% at 2\,Myr and 2.8\% at 10\,Myr. This is similar to \citet{Fedele10_accretion_timescale} which reports 60\% and 2\%, respectively. We detect 22 very old accretors (in regions with age $>10$\,Myr), which is 3.7\% of the total number of accretors in Sco-Cen.

In addition to the fraction of accretors at different ages, we analysed the evolution of the median accretion luminosity and mass accretion rate of the accretors in each cluster as a function of cluster age. To ensure that we were tracing real accretion, we again only considered the sample of sources that have pEW\,$<-1.0$\,nm. For this analysis, we used the values calculated with med-GSP-Phot extinction, because while values calculated with GSP-Phot extinction better correlate with literature values (Sect. \ref{sec: lit}) they are only available for 28\% of the Sco-Cen sample. Therefore, to consider a large enough number of sources per cluster for running a statistical analysis, we chose to use med-GSP-Phot extinction corrected values (available for 98\% of the sample), keeping in mind that they represent a consistent lower limit to the literature values.

Before proceeding with the analysis, we first had to evaluate any dependence with stellar mass, as it strongly correlates with accretion (Fig. \ref{fig:Lstar_Lacc_Macc_relation}). Using the stellar parameters derived in Sect. \ref{sec:derivation_stellar_parameters} (Table \ref{table:AccretionTable_extract} from med-GSP-Phot extinctions, we observe differences between the median stellar mass of each of the Sco-Cen clusters (Fig. \ref{fig:Mass_distribution_scocen}). In particular, we observe significant higher median stellar masses in Chameleon I and II. This difference is also present if the stellar masses derived from spectroscopy from \citet{Manara_TTauri} are considered. We believe this is due to the fact that Chameleon I and II are some of the most Southern regions, making the populations harder to observe for most observatories. This is particularly relevant for faint sources, causing their observed populations to have higher median mass values than other regions. To avoid tracing the effect of stellar mass on accretion, we only considered the sources with stellar mass between $0.08$ M$_\odot$ and $0.25$ M$_\odot$ in every cluster. This forces the median stellar mass of all clusters to be similar. Then, for each cluster where there are at least four accretors in the chosen mass range, we calculated the median accretion luminosity and mass accretion rate. Uncertainties are given by the 16\textsuperscript{th} and 84\textsuperscript{84} percentiles. Values of the medians, both corrected and uncorrected for stellar mass, can be found in Table \ref{table:ScoCen}. The evolution of the median accretion luminosity and mass accretion rate with age are shown in the bottom two panels of Fig. \ref{fig:timescales_scocen}. We again fit an exponential decay function and a power law function, of the same form as the ones shown, respectively, in eqns. \ref{eqn:timescale_fraction} and \ref{eqn:power_law}. The results of these fits, performed with the approach from \citet{linmix} to account for errors, are reported in Table \ref{Table:scocen_parameters}. As for the case of $f_\text{acc}$, the power law appears to better fit the data. The timescales, derived from the exponential fit, are larger than the one found for $f_\text{acc}$, but they also have large uncertainties. Comparing the values of $\alpha$, the exponent of the power law, also shows that the evolution of median $L_{\text{acc}}$ and median $\dot{M}_\text{acc}$ with time is slower than that of $f_\text{acc}$.

\begin{table*}[ht]
    \caption{Fit parameters for the evolution of fraction of accretors ($f_{\text{acc}}$), median $L_\text{acc}$ of the accretors and median $\dot{M}_\text{acc}$ of the accretors in Sco-Cen clusters with age, where by accretors we mean sources with pEW\,$<-1.0$\,nm.}
    \centering

    \small
    \begin{tabular}{l c l c c c l c}
    \hline\hline\\[-0.9em]   
     & \multicolumn{3}{c}{Exponential Decay} & &  \multicolumn{3}{c}{Power Law} \\   
    \cline{2-4}  
    \cline{6-8}
    \\[-0.8em] 
       $y$ & $\tau_{\text{acc}}$ [Myr] & $C$ & $\chi_\text{red}^2$ & & $\alpha$ & $k$ & $\chi_\text{red}^2$\\
    \noalign{\smallskip}
    \hline
    \noalign{\smallskip}
    $f_{\text{acc}}$ (pEW\,$<-1$) & $2.7 \pm 0.4$ & $79 \pm 28$ & 6.1 & & $-2.0 \pm 0.3$ & $281 \pm 149$ & 3.4\\
    $L_\text{acc}$ & $4.9 \pm 3.0$ & $4.7 \pm 4.2$ [$10^{-4} L_\odot$] & 1.2 & & $-1.3 \pm 0.6$ & $1.3 \pm 1.7$ [$10^{-3} L_\odot$] & 0.46 \\
    $\dot{M}_\text{acc}$ & $4.5 \pm 2.4$ & $7.6 \pm 5.5$ [$10^{-11} M_\odot$/yr] & 0.60 & & $-1.5 \pm 0.6$ & $3.1 \pm 3.8$ [$10^{-10} M_\odot$/yr] & 0.39 \\
    \hline
    \end{tabular}

    \tablefoot{ For each quantity, two fits are preformed: an exponential decay of the form $y = C\cdot \exp{(-t/\tau_{\text{acc}})}$, and a power law of the form $y = k\cdot t^\alpha$. Here, $t$ is the age in Myr. Reduced $\chi^2$ values are also reported for each fit.}
    
    \label{Table:scocen_parameters}
\end{table*}

\citet{2024arXiv241021598C} compared millimetre-wavelength observations with the \citet{ScoCen_SigMA_ages,ScoCen_SigMA_sources} cluster membership in Upper Scorpius, and found no significant evolution of mm-fluxes with age. Indeed, \citet{Polnitzky2025} finds a significantly larger than 2.7 Myr disc dissipation timescale for the same Sco-Cen clusters considered in this work. This could suggest the presence of planet-forming discs containing dust-traps (e.g. \citealp{2022A&A...663A..98T}), for which significant primordial or second-generation dust-disc material could be still detected while accretion is below detection thresholds. Indeed, a larger fraction of discs with inner dust cavities has been tentatively reported at the later Upper Scorpius ages \citep{Vioque2025_acc}.

A tentative steepening with time of the $M_{\text{disc}}$-$M_{\star}$ relation has been proposed (e.g. \citealp{2016ApJ...831..125P,2017AJ....153..240A}), although this steepening is contested due to the scatter in the data and the associated uncertainties (\citealp{2022A&A...663A..98T, Manara_TTauri}). If this steepening is true, population synthesis models have predicted that for viscously evolving discs the $\dot{M}_{\text{acc}}$-$M_{\star}$ correlation (Fig. \ref{fig:Lstar_Lacc_Macc_relation}) should also steepen with time (\citealp{2022MNRAS.514.5927S}). This would imply that the accretion timescale for low mass YSOs is shorter (i.e. they evolve faster) than for more massive YSOs. By taking the Sco-Cen clusters of \citet{ScoCen_SigMA_ages,ScoCen_SigMA_sources} we can evaluate this change in steepness in the $\dot{M}_{\text{acc}}$-$M_{\star}$ and $L_{\text{acc}}$-$L_{\star}$ power law relationships homogeneously for the whole Sco-Cen region (spanning $\sim10$ Myr). The results of fitting the two relations for the ten clusters of Sco-Cen with at least 20 sources for which we traced accretion (considering only the sources with pEW\,$<-1.0$\,nm to exclude contaminants, as in criterion B of Sect. \ref{Sect:YSO_flags_criteria}) are presented in Fig. \ref{figure_slope_change}. The plots of the fits for each region, and the values of the slopes, can be found in Appendix \ref{Appendix scocen}. For most clusters, the slopes we find are in the same range as those calculated in \citet{Almendros-Abad24}, which are discussed in Sect. \ref{Sec:Accretion_all_sky}. The correlations we find are not indicating any clear evolution with time of either the $\dot{M}_{\text{acc}}$-$M_{\star}$ or the $L_{\text{acc}}$-$L_{\star}$ relations. We do not find significant gradient differences for clusters of very different ages. We note that in this analysis we do not apply any filtering on the mass range of cluster members. More accurate determinations of accretion rates for complete populations are needed to further explore these relations.

\begin{figure}[h]
    \centering
    \includegraphics[width=0.95\linewidth]{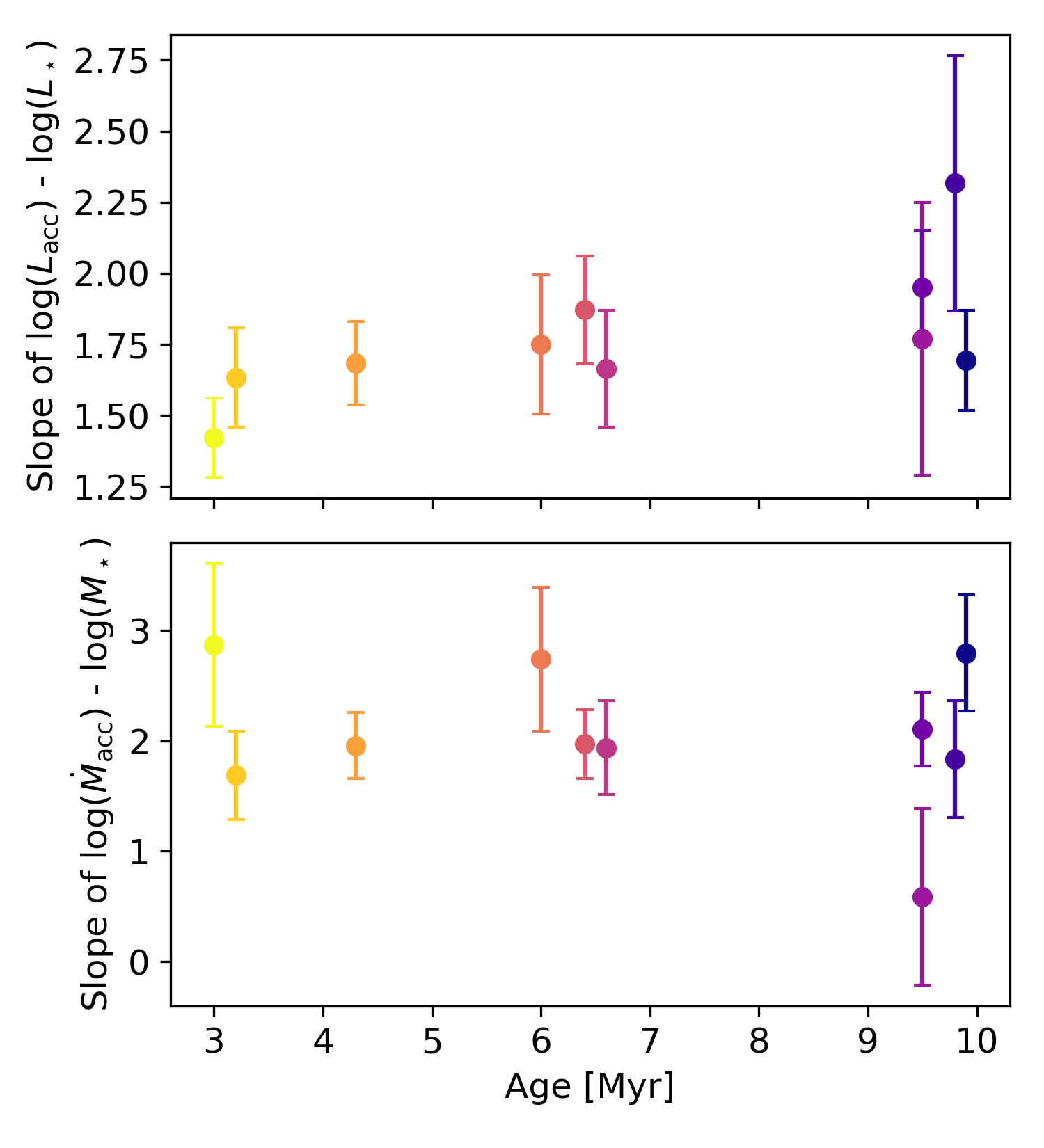}
    \caption{Slope of the correlations of accretion luminosity vs stellar luminosity (top panel), and mass accretion rate vs stellar mass (bottom panel) for ten clusters of Sco-Cen, plotted against cluster age (\citealp{BHAC15} Bp-Rp ages). For each cluster, we measured the slope of the correlations in the logarithmic plane, using values calculated with med-GSP-Phot, and only considering sources with pEW\,$<-1.0$\,nm. Only clusters where at least 20 sources could be used in the fit were considered. Each point is colour-coded to match the corresponding fit in Appendix \ref{Appendix scocen}.} \label{figure_slope_change}
\end{figure}

\subsection{Other star-forming regions}\label{Sec_other_regions}

An implicit assumption of Sect. \ref{sec:ScoCen_ages} is that all clusters within Sco-Cen can be considered as snapshots of the same star formation process, disregarding the impact of the environment or cluster-to-cluster intrinsic differences. While this assumption might be appropriate for the general derivations of Sect. \ref{sec:ScoCen_ages} for the Sco-Cen complex, it can help explain some of the observed scatter. For example, external photoevaporation from massive stars has been proven to have a significant impact in Upper Scorpius gas-disc radii \citep{Anania2025}.

For legacy value, in this section we derive the fraction of accretors ($f_{\text{acc}}$), median accretion luminosity and median mass accretion rate for other known star-forming regions and groups (following the prescriptions of Sect. \ref{sec:ScoCen_ages} and using Table \ref{table:AccretionTable_extract}). We report the derived values in Table \ref{tab:df_pEW_i}, together with literature estimates of ages and the references for group membership. 

\begin{figure}[h]
    \centering
    \includegraphics[width=\linewidth]{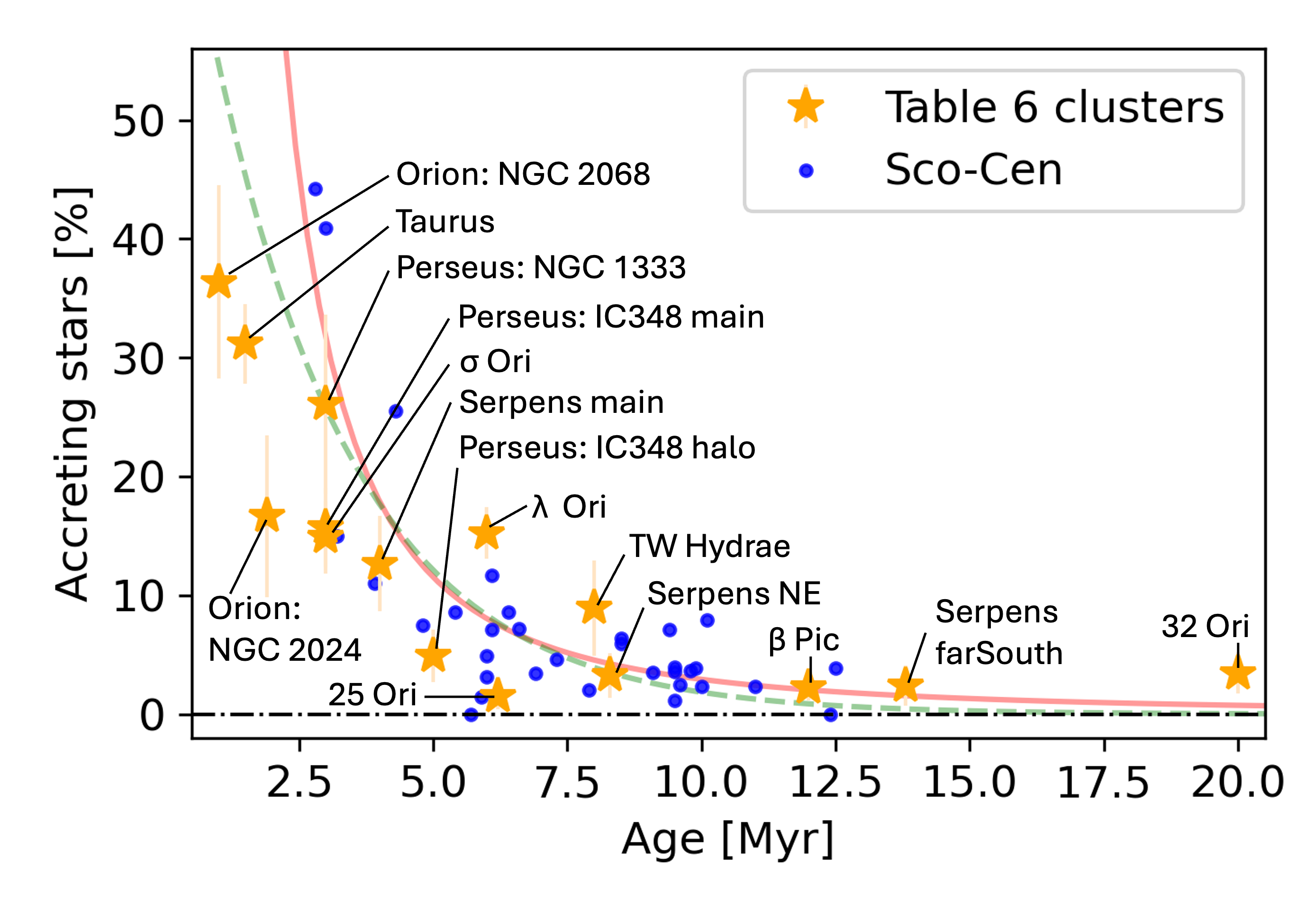}
    \caption{Fraction of accretors against age for each region of Table \ref{tab:df_pEW_i}, over-plotted on the top panel of Fig. \ref{fig:timescales_scocen}. The Table \ref{tab:df_pEW_i} regions (orange stars in the plot) are labelled. The blue dots are the Sco-Cen clusters of Fig. \ref{fig:timescales_scocen} and the red and green line are, respectively, the power law and exponential fit to the Sco-Cen clusters, obtained in Sect. \ref{sec:ScoCen_ages}.}
    \label{fig:table6_figure}
\end{figure}

Table \ref{tab:df_pEW_i} shows that older regions tend to have lower values of median accretion luminosity and mass accretion rate. The fraction of accretors ($f_{\text{acc}}$) per cluster also decreases with time, as illustrated in Fig. \ref{fig:table6_figure}, where the regions of Table \ref{tab:df_pEW_i} are over-plotted on the top panel of Fig. \ref{fig:timescales_scocen}. We note that the regions of Table \ref{tab:df_pEW_i} span very different environments and their ages come from different sources. In addition, they vary in distance and extinction and hence \textit{Gaia} traces different YSO populations in them. Thus, Fig. \ref{fig:table6_figure} is only illustrative.

\begin{table*}
 \caption[]{\label{tab:df_pEW_i}Fraction of accreting stars, median accretion luminosity, and median mass accretion rate for various star-forming regions and groups within 500\,pc.}
\resizebox{\textwidth}{!}{\begin{tabular}{l c c c c c c c }
\hline\hline\\[-0.9em]   
& & \multicolumn{2}{c}{References} & & & \multicolumn{2}{c}{Median values per region}\\
    \cline{3-4}
    \cline{7-8}\\[-0.8em] 
Region & Age [Myr] & Membership & Age & N. with pEW & Accretors [\%] & $L_{\text{acc}}$  [$L_\odot$] & $\dot{M}_\text{acc}$ [$M_\odot$ yr$^{-1}$]\\
\hline\\[-0.8em] 
\textit{Orion} NGC 2068       & 1.0  & K18 & K18 & 55  & $36 \pm 8$ & 8.26E$-3$ (2.35E$-3$, 2.53E$-2$) & 8.79E$-10$ (3.26E$-10$, 2.22E$-09$) \\ 
Taurus          & 1.5  & L18 & F10 & 279 & $31 \pm 3 $& 1.98E$-3$ (2.97E$-4$, 4.70E$-2$) & 3.01E$-10$ (4.53E$-11$, 3.48E$-09$)   \\ 
\textit{Orion} NGC 2024       & 1.9  & Z24 & K18 & 36  & $17 \pm 7$ & 1.38E$-3$ (9.26E$-4$, 4.10E$-3$) & 2.01E$-10$ (8.43E$-11$, 5.14E$-10$)  \\ 
\textit{Perseus} IC348 main & 3.0 & O23 & O23 & 140 & $16 \pm 3$ & 5.89E$-3$ (6.58E$-4$, 2.10E$-2$) & 5.11E$-10$ (9.69E$-11$, 1.85E$-09$)   \\ 
\textit{Perseus} NGC1333  & 3.0  & O23 & O23 & 46  & $26 \pm 8$ & 5.10E$-3$ (2.03E$-3$, 2.17E$-2$) & 4.95E$-10$ (1.85E$-10$, 1.51E$-09$)  \\ 
\textit{Orion} $\sigma$ Orionis      & 3.0  & Z24 & K18 & 161 & $15 \pm 3$ & 1.75E$-3$ (6.50E$-4$, 9.27E$-3$) & 2.56E$-10$ (6.11E$-11$, 7.04E$-10$) \\ 
\textit{Serpens} main   & 4.0  & H19 & Z22 & 79  & $13 \pm 4$ & 2.91E$-2$ (2.41E$-3$, 1.75E$-1$) & 2.18E$-09$ (3.27E$-10$, 1.24E$-08$)\\ 
\textit{Perseus} IC348 halo & 5.0 & O23 & O23 & 102 & $16 \pm 3$  & 2.22E$-3$ (7.93E$-4$, 1.23E$-2$) & 3.46E$-10$ (6.11E$-11$, 8.01E$-10$) \\ 
\textit{Orion} $\lambda$ Orionis & 6.0  & C21 & K18 & 328 & $15 \pm 2$ & 8.48E$-3$ (2.51E$-3$, 4.59E$-2$) & 7.94E$-10$ (1.93E$-10$, 2.76E$-09$) \\ 
\textit{Orion} 25 Orionis & 6.2  & K18 & K18 & 134 & $1.5 \pm 1$  &  &  \\
TW Hydrae         & 8.0  & L23 & F10 & 56  & $9 \pm 4$  & 2.41E$-4$ (1.09E$-5$, 2.60E$-2$) & 1.56E$-11$ (1.76E$-12$, 1.32E$-09$)  \\ 
\textit{Serpens} NorthEast     & 8.3  & H19 & Z22 & 92  & $3 \pm 2$  &  &  \\ 
$\beta$ Pictoris       & 12.0 & L24 & F10 & 177 & $2 \pm 1$  & 5.55E$-6$ (2.54E$-6$, 2.53E$-5$) & 9.08E$-13$ (4.01E$-13$, 5.55E$-12$)  \\
\textit{Serpens} farSouth & 13.8 & H19 & Z22 & 84  & $2.4 \pm 1.7$  &  &  \\ 
\textit{Orion} 32 Orionis         & 20.0 & L22 & L22 & 116 & $3 \pm 2$  & 4.31E$-5$ (1.97E$-5$, 3.65E$-4$) & 3.84E$-12$ (3.30E$-12$, 2.71E$-11$)  \\ 

\hline

\end{tabular}}

\tablefoot{Median values reported here are calculated considering sources with pEW\,$<-1.0$\,nm and using the med-GSP-Phot extinction correction, for direct comparison with Table \ref{table:ScoCen}. Median values are only reported for groups where four or more sources could be used for the calculation. The 16\textsuperscript{th} and 84\textsuperscript{th} percentile are in brackets. The full table is available at the CDS and also provides the values calculated with GSP-Phot extinction.}

\tablebib{(C21) \citet{C21}; (F10) \citet{Fedele10_accretion_timescale}; (H19) \citet{sf_H19};
(K18) \citet{sf_Orion_kounkel18}; (L18) \citet{sf_Taurus_Luhman18}; (L22) \citet{sf_L22};
(L23) \citet{sf_TWA_Luhman23}; (L24) \citet{sf_BetaPic_Lee24}; (O23) \citet{sf_O23}; (Z22) \citet{sf_Z22};
(Z24) \citet{sf_Z24}.}
\end{table*}

\section{Conclusions} \label{Sect:conclusions}

We have presented an all-sky homogeneous analysis of the H\textalpha{} emitting YSO candidates within 500\,pc down to G magnitude 17.65\,mag. By using the \textit{Gaia} XP spectra, we have characterised the H\textalpha{} line and derived accretion luminosities and mass accretion rates for a sample of $145\,975$ potential YSOs, which we present in Table \ref{table:AccretionTable_extract}, together with an estimation of the stellar parameters. We evidence the accuracy of our accretion determinations by comparing to accretion measurements derived from higher-resolution spectra. In Sect. \ref{Sect:YSO_flags_criteria} we describe filtering strategies to select specific sub-samples of YSOs from Table \ref{table:AccretionTable_extract} that could be adopted for specific science cases.

We present some applications of this YSO catalogue for studying star formation in the solar neighbourhood. The main results we obtain after analysing this new sample of YSOs are:

\begin{enumerate}
    \item While we recover well the known population of YSOs with measured accretion rates, we identify a large population of low-accreting YSO candidates previously untraced by YSO accretion rate surveys. Even when applying strict sample-purity constraints, we recover a significant population of accreting YSOs at $\dot{M}_{\text{acc}}=10^{-11}$ to $10^{-12}$ $M_{\odot}$/yr. We find previous accretion rate studies have mostly focused on the YSO population with significant infrared excesses from disc emission.
    
    \item The population of low-accreting potential YSOs is mostly dispersed, away from star-forming regions or clustered environments of star formation. Even within star-forming regions, the YSO low accretors appear disperse and separated from the more clustered areas, where the accretion luminosities and mass accretion rates are higher. This could be due to the influence of the environment on star formation or YSO dispersion with age. Many low accretors appear entirely disconnected from young regions. This disperse population is likely to contain new `Peter Pan' YSOs.

    \item We evaluated the $\dot{M}_{\text{acc}}$-$M_{\star}$ and $L_{\text{acc}}$-$L_{\star}$ relations for our sample of all-sky YSOs, finding values of $L_{\text{acc}}\propto L_\star ^{1.41\pm 0.02}$ and $\dot{M}_\text{acc} \propto M_{\star}^{2.4 \pm 0.1}$. Both correlations are broadly consistent with literature values. In addition, we report the values of this relation for regions of different ages across the Sco-Cen complex, finding that the $\dot{M}_{\text{acc}}$-M$_{\star}$ and $L_{\text{acc}}$-$L_{\star}$ correlations remain roughly constant in time.

    \item We analysed the decay of accretion rate with time by using the Sco-Cen complex cluster members and ages of \citet[Table \ref{table:ScoCen}]{ScoCen_SigMA_ages,ScoCen_SigMA_sources}. By fitting an exponential function to the fraction of accreting stars in clusters of different ages, we obtain an accretion timescale of $\tau_{\text{acc}} = 2.7 \pm 0.4$\,Myr. We find that a power law fits the evolution with age better than an exponential function and, using a power law fit, we find that the percentage of accretors is $70\%$ at 2\,Myr and 2.8\% at $10$\,Myr.  Moreover, we fit the exponential decay timescales of $L_{\text{acc}}$ ($\tau_{\text{acc}} = 4.9 \pm 3.0$\,Myr) and $\dot M_{\text{acc}}$ ($\tau_{\text{acc}} = 4.5 \pm 2.4$\,Myr) of accreting sources. For all three cases, we report the results of fitting both the exponential function and the power law.

    \item We report the fraction of accreting stars, median accretion luminosity and median mass accretion rate, for a collection of star-forming regions and groups within 500\,pc (Table \ref{tab:df_pEW_i}). We confirm the decay of accretion with age in these regions.
    
\end{enumerate}

This work constitutes the first effort to derive accretion rates for YSOs using the \textit{Gaia} XP spectra. We have produced a sample of 145\,975 nearby accreting YSO candidates, with 4\,208 being more robust based on their IR excess, 6\,170 being more robust based on their H\textalpha{} emission strength, and 1\,945 sources being the most robust, based on both factors. This work lays the foundations for improved analyses to be done with \textit{Gaia} data release 4, and the publication of all the individual epochs of XP spectra. Other ongoing and future efforts to characterise in greater detail the extinction towards the different star-forming regions will also benefit the accuracy of future derivations.

\section*{Data availability}
Full Tables \ref{table:AccretionTable_extract}, \ref{table:ScoCen} and \ref{tab:df_pEW_i} are only available in electronic form at the CDS via anonymous ftp to \url{cdsarc.u-strasbg.fr} (130.79.128.5) or via \url{https://cdsarc.cds.unistra.fr/viz-bin/cat/J/A+A/699/A145}.

\begin{acknowledgements}
We thank the anonymous referee for their valuable comments and suggestions which improved the quality of the paper. We thank Giacomo Beccari, Carlo F. Manara, and Alice Somigliana for insightful discussions which improved this work. This job has made use of the Python package GaiaXPy, developed and maintained by members of the \textit{Gaia} Data Processing and Analysis Consortium (DPAC), and in particular, Coordination Unit 5 (CU5), and the Data Processing Centre located at the Institute of Astronomy, Cambridge, UK (DPCI). 
This work has made use of data from the European Space Agency (ESA) mission
{\it Gaia} (\url{https://www.cosmos.esa.int/gaia}), processed by the {\it Gaia}
Data Processing and Analysis Consortium (DPAC,
\url{https://www.cosmos.esa.int/web/gaia/dpac/consortium}). Funding for the DPAC
has been provided by national institutions, in particular the institutions
participating in the {\it Gaia} Multilateral Agreement. This research has made use of the TOPCAT interactive graphical tool (\citealp{2005ASPC..347...29T}). Additionally, this research has made use of the VizieR catalogue access tool, CDS,
Strasbourg, France.
\end{acknowledgements}

\bibliographystyle{aa}
\bibliography{bibliography}

\begin{appendix}
\section{Sample query}\label{Appendix A}

ADQL query used to select the sample of all sources compatible with distance $<500$\,pc, for which H\textalpha{} pseudo-equivalent width measurements (pEW) are available in the \textit{Gaia} DR3 table of astrophysical parameters (\texttt{astrophysical\_parameters}, \citealp{ast_par_creevy}).

\begin{lstlisting}
SELECT *
FROM gaiadr3.astrophysical_parameters, 
external.gaiaedr3_distance, gaiadr3.gaia_source
WHERE ew_espels_halpha IS NOT NULL
AND gaia_source.source_id = gaiaedr3_distance.source_id 
AND gaia_source.source_id = astrophysical_parameters.source_id
AND gaiaedr3_distance.r_lo_geo<500
\end{lstlisting}

\section{Identifying non-H-alpha emitter contaminants} \label{Appendix B}

A number of sources with no H\textalpha{} emission can appear to have it in the low-resolution \textit{Gaia} XP spectra. These contaminants are mostly M-dwarfs. M-dwarfs are cool low mass stars that comprise around 75\% of stars in the Milky Way \citep{Mdwarfs_EDR3}. At low temperatures, titanium oxide (TiO) is present in the atmosphere of M-dwarfs and causes absorption bands in the spectra \citep{M-dwarfs}, including in the region around the H\textalpha{} emission wavelength \citep{M_dwarfs_2}. In high-resolution spectra the TiO absorption lines are resolved and various extrema can be observed in the H\textalpha{} region. However, in low-resolution spectra the individual spectral features are not well resolved and the TiO absorption bands appear as a broad feature around 656.3\,nm, which \texttt{linefinder} mistakes for a H\textalpha{} emission line. For stars with higher effective temperatures than M-dwarfs TiO dissociates and the absorption bands are not present. Since most M-dwarfs are not true H\textalpha{} emitters, in this appendix we present a method for excluding M-dwarfs from the sample of candidate H\textalpha{} emitters.

\begin{figure*}
    \centering
    \vspace{2\baselineskip}
    \includegraphics[width=\linewidth]{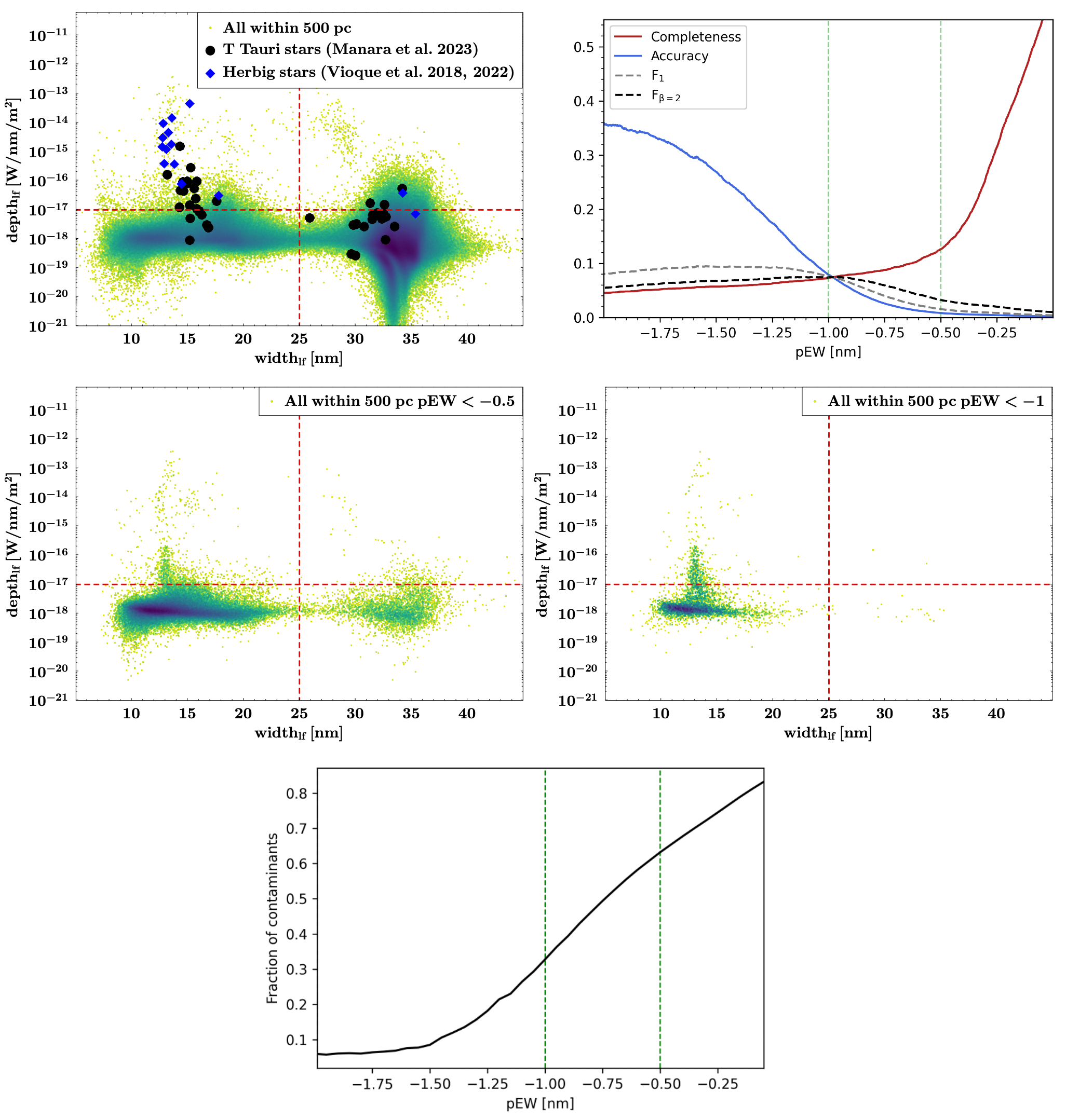}
    \caption{\textit{Top left:} Density plot of the H\textalpha{} line-depth against line-width as identified by \texttt{linefinder} for all sources within 500\,pc with \textit{Gaia} DR3 XP spectra available. Dashed red lines indicate the thresholds in line-width and line-depth used to filter non-real H\textalpha{} emitters (see Appendix \ref{Appendix B}). Sources with \textit{width\textsubscript{lf}}\,$>25$ nm are mostly M-dwarf contaminants. Sources with \textit{depth\textsubscript{lf}} $<10^{-17}$\,W/nm/m$^2$ have emission comparable to the \texttt{linefinder} uncertainties. \textit{Top right:} `accuracy', `completeness', `F\textsubscript{1}-score', and `F\textsubscript{$\rm \beta$}-score' metrics as a function of pEW threshold for selecting H\textalpha{} emitters. Vertical lines indicate the main thresholds used in this work (pEW\,$<-0.5$, and $<-1.0$\,nm). \textit{Middle panels:} Density plots of the H\textalpha{} line-depth against line-width as identified by \texttt{linefinder} for all sources within 500\,pc with \textit{Gaia} DR3 XP spectra available and pEW\,$<-0.5$\,nm (left) or pEW\,$<-1.0$\,nm (right). \textit{Bottom panel:} Plot of the fraction of contaminants against pEW threshold applied, where the number of contaminants is defined as the number of sources with with \textit{width\textsubscript{lf}} $>25$\,nm.}
    \label{fig:depth-width}
\end{figure*}

According to \texttt{linefinder}, there are $2\,756\,293$ sources with H\textalpha{} emission within 500\,pc.  We started by assessing the line-parameters \textit{depth\textsubscript{lf}} and \textit{width\textsubscript{lf}} which \texttt{linefinder} reports for this sample (top left panel of Fig. \ref{fig:depth-width}). Since TiO causes a broad line in the H\textalpha{} region of the low-resolution spectra, the measured `H\textalpha{} line-width' can be a good tracer for identifying M-dwarf contaminants. Indeed, in Fig. \ref{fig:depth-width} a clear division into two populations of different \textit{width\textsubscript{lf}} can be appreciated: one population with narrow lines, as expected for real H\textalpha{} emission, and one population with wide lines, as expected for M-dwarf contaminants. The dividing threshold is at around \textit{width\textsubscript{lf}} $=25$ nm. We confirmed that sources with \textit{width\textsubscript{lf}} $>25$ nm are predominantly M-dwarfs by checking their location on a colour-magnitude diagram and visually inspecting their spectra. Therefore, we consider all sources with width$_{lf}> 25$\,nm as M-dwarf contaminants with no real H\textalpha{} emission as traced by \texttt{linefinder} from the \textit{Gaia} XP spectra, and exclude them from the analyses of this work (see Fig. \ref{fig:selection_diagram}). We note that M-dwarfs with true H\textalpha{} emission at the level of the TiO feature are also excluded by this cut.

It is also necessary to consider where is the noise level in the H\textalpha{} \texttt{linefinder} line detection. Some \textit{depth\textsubscript{lf}} are so small it is unclear if they are tracing true emission (Fig. \ref{fig:depth-width}). In order to evaluate this, we ran \texttt{linefinder} on well-known YSOs with H\textalpha{} emission from the works of \citet[T Tauri stars]{Manara_TTauri} and \citet[Herbig stars]{Vioque_2018,Vioque_2022} that have public XP spectra (40 T Tauris and 14 Herbigs in total). We found that, of those sources with \textit{width\textsubscript{lf}} $<25$ nm, \textit{depth\textsubscript{lf}} $>10^{-17}$\,W/nm/m$^2$ is a good threshold for tracing real H\textalpha{} emission (top left panel of Fig. \ref{fig:depth-width}). Lower values of \textit{depth\textsubscript{lf}} include a significant number of sources and very few of the well-known YSOs. This latter sample selection cut of \textit{depth\textsubscript{lf}}$>10^{-17}$\,W/nm/m$^2$ can be used in cases where it is acceptable to only consider the strongest H\textalpha{} emitters, which are less likely to be contaminants. However, as well as removing false positives it also removes real faint H\textalpha{} emitters.

As described in Sect. \ref{Sect:sample_selection}, not all sources with pEW have published XP spectra (although the inverse is true), and this includes the majority of historically known YSOs. It is thus necessary to select a threshold in pEW that selects true H\textalpha{} emitters (i.e. maximises `accuracy'), while retaining the largest number of true emitters as possible (i.e. maximises `completeness'). This can be done via the F\textsubscript{1}-score, a harmonic mean of accuracy and completeness. However, in this work it is more interesting to prioritise completeness, as we have independent means to identify contaminants (Fig. \ref{fig:selection_diagram}). Hence, we use a more general F\textsubscript{$\rm \beta$}-score, with $\rm \beta=2$, which weights completeness higher than accuracy. To define these metrics in a \textit{Gaia} XP self-consistent manner, we consider as true H\textalpha{} emitters all objects within 500\,pc with \textit{width\textsubscript{lf}} $<25$ nm and \textit{depth\textsubscript{lf}} $>10^{-17}$\,W/nm/m$^2$, and as false H\textalpha{} emitters the rest of sources in that parameter space (top left panel of Fig. \ref{fig:depth-width}). The fraction of false H\textalpha{} emitters as a function of adopted pEW selection threshold is shown in the bottom panel of Fig. \ref{fig:depth-width}, showing the fraction of contaminants decreasing as the threshold becomes more conservative. We can then build the `accuracy', `completeness', `F\textsubscript{1}-score', and `F\textsubscript{$\rm \beta$}-score' metrics as a function of pEW selection threshold. These metrics are shown on the top right panel of Fig. \ref{fig:depth-width} as a function of pEW threshold for selecting H\textalpha{} emitters. The `F\textsubscript{$\rm \beta$}-score' is maximum when selecting objects with pEW\,$<-1.0$\,nm, and hence we consider this value as the best threshold for selecting true H\textalpha{} emitters. However, pEW\,$<-1.0$\,nm drastically reduce the sample size from $2\;756\;293$ to $5\;116$ sources. A more complete threshold is to consider sources with pEW\,$<-0.5$\,nm, which reduces the accuracy of the selection significantly but greatly increases the completeness ($95\;009$ sources) and the bias against weak H\textalpha{} emitters. Independent observables can be then used to further filter YSO-like objects. Therefore, different pEW thresholds can be used depending on the use case to select samples of H\textalpha{} emitting sources. The resulting populations in the \textit{width\textsubscript{lf}} and \textit{depth\textsubscript{lf}} parameter space after filtering by pEW $<-1.0$, and $<-0.5$ nm are shown in the middle panels of Fig. \ref{fig:depth-width}.

The results of running \texttt{linefinder} with \texttt{truncation = false} on all sources within 500\,pc with pEW\,$<-0.5$\,nm are included in Table \ref{table:AccretionTable_extract}. Of $150\;383$ total sources, $130\;109$ have XP spectra available and of these, \texttt{linefinder} detects H\textalpha{} for $95\;094$. Users looking for purer samples of true strong H\textalpha{} emitters might want to use more conservative thresholds of pEW (e.g. pEW\,$<-1.0$\,nm) or use the recommended \textit{width\textsubscript{lf}} $<25$ nm and \textit{depth\textsubscript{lf}} $>10^{-17}$\,W/nm/m$^2$ if XP spectra is available.

\vspace{6\baselineskip}

\section{Correlation between $F_\text{cont}$ and RP}\label{Appendix:Fcont_RP_correlation}

\begin{figure}[h!]
    \centering
    \includegraphics[width=\linewidth]{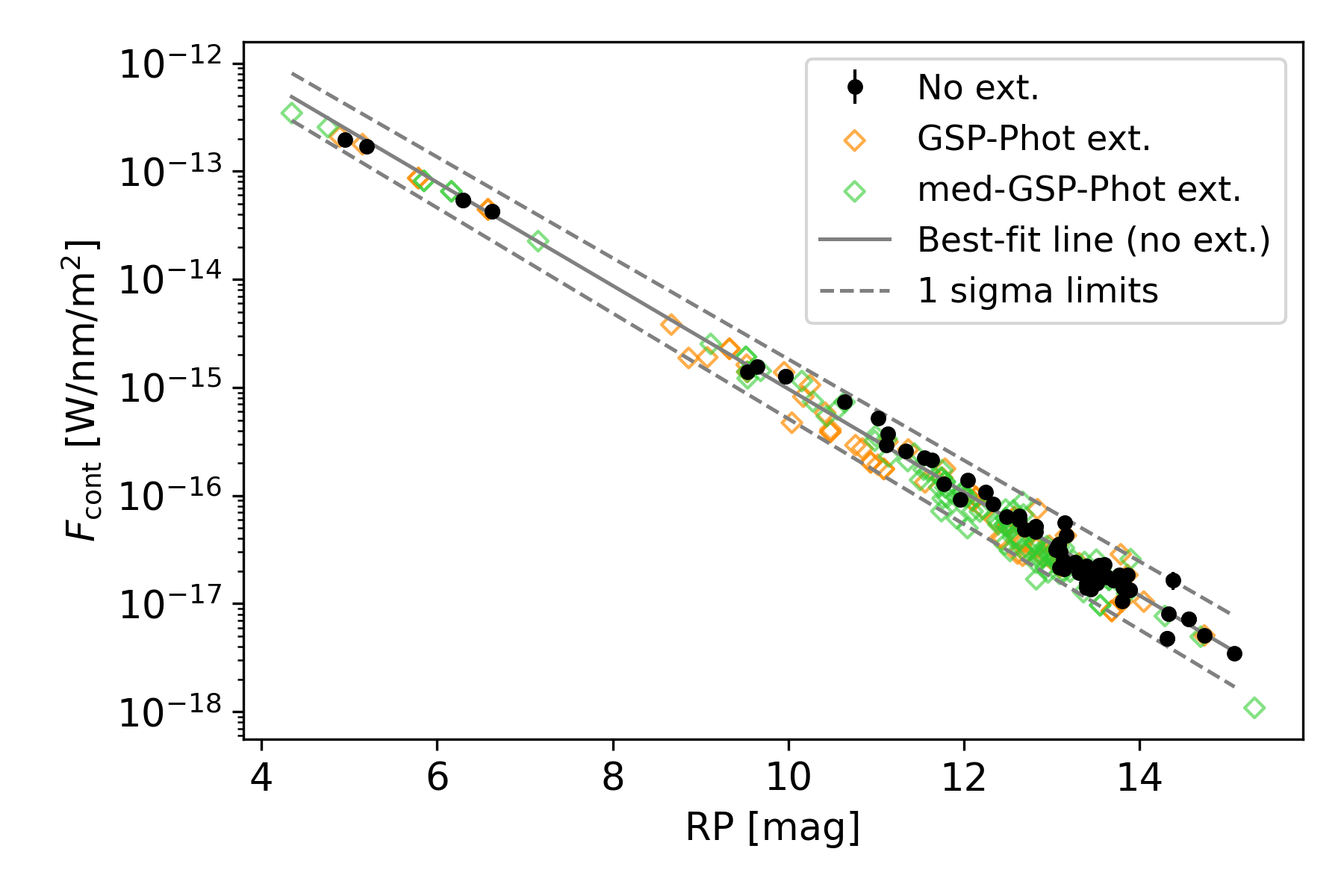}
    \caption{Correlation between the continuum flux density at the H\textalpha{} line ($F_{\text{cont}}$) and the mean magnitude in the integrated \textit{Gaia} RP band, for a sample of YSOs for which $F_{\text{cont}}$ was calculated from the XP spectra. The points show all three extinction regimes discussed in Sect. \ref{ext_section}. The line of best linear fit, with its uncertainties, is only shown for the case of no extinction correction. The values of the fit parameters for all three cases are presented in Table \ref{table:Fcont_RP_coefficients}.}
    \label{fig:Fcont_RP_correlation}
\end{figure}

Fig. \ref{fig:Fcont_RP_correlation} shows the correlation between $F_\text{cont}$ and RP for the three different extinction regimes discussed in Sect. \ref{ext_section}.

\section{YSO region on colour-magnitude diagram}\label{Appendix C}

To remove non-YSOs from the catalogue of Table \ref{table:AccretionTable_extract}, we apply a selection criterion based on the \textit{Gaia} colour-magnitude diagram (BP-RP vs absolute G magnitude). We convert the G band mean magnitude to absolute magnitude using the geometric distances of \citet{geom_dist}. As the goal is to perform a filtering of obvious non-YSO objects, we focus on highlighting the regions clearly incompatible with YSOs in the CMD (by adding generous margins to the \citealp{BHAC15} tracks). In particular, we select as YSO-compatible the sources that fall in the area bounded by the following lines (Fig. \ref{fig:HR_cut}):
\begin{equation}
    y=-2,\;\;y=2.67x-10.01,\;\;y=4.17x+3.82,\;\;y=-2.
\end{equation}

This CMD cut excludes populations such as white dwarfs, cataclysmic variables, evolved stars, and Classical Be stars. Sources within the YSO-compatible region are flagged in Table \ref{table:AccretionTable_extract} as `flag\_CMD'.

\section{Addressing the inhomogeneous sky distribution of sources with no XP spectra} \label{Appendix D}

The sky distribution of sources that have no XP spectra in \textit{Gaia} DR3, but that have pEW measurements available in the \texttt{astrophysical\_parameters} table, is affected by the \textit{Gaia} scanning law. \textit{Gaia} scans some regions of the sky less frequently than others \citep{gaia_mission}, and XP spectra was not published in DR3 for sources that have fewer than 15 CCD transits contributing to the generation of the BP and RP spectra \citep{xp_DeAngeli}. However, pEWs were published for these sources regardless of them not having public XP spectra. We thus need to account for the fact that some entire regions of the sky have no XP spectra, and hence there is no possible removal of M-dwarf contaminants via \texttt{linefinder} (see Sect. \ref{sec:Lacc_derivation} and Appendix \ref{Appendix B}). The obvious approach of not considering pEWs for sources with no XP spectra is not adequate. This is because, importantly, most known YSOs have no public XP spectra due to unrelated internal reasons (private communication with \textit{Gaia} helpdesk), but they also have pEW measurements in DR3. 

\begin{figure}
    \centering
    \includegraphics[width=\linewidth]{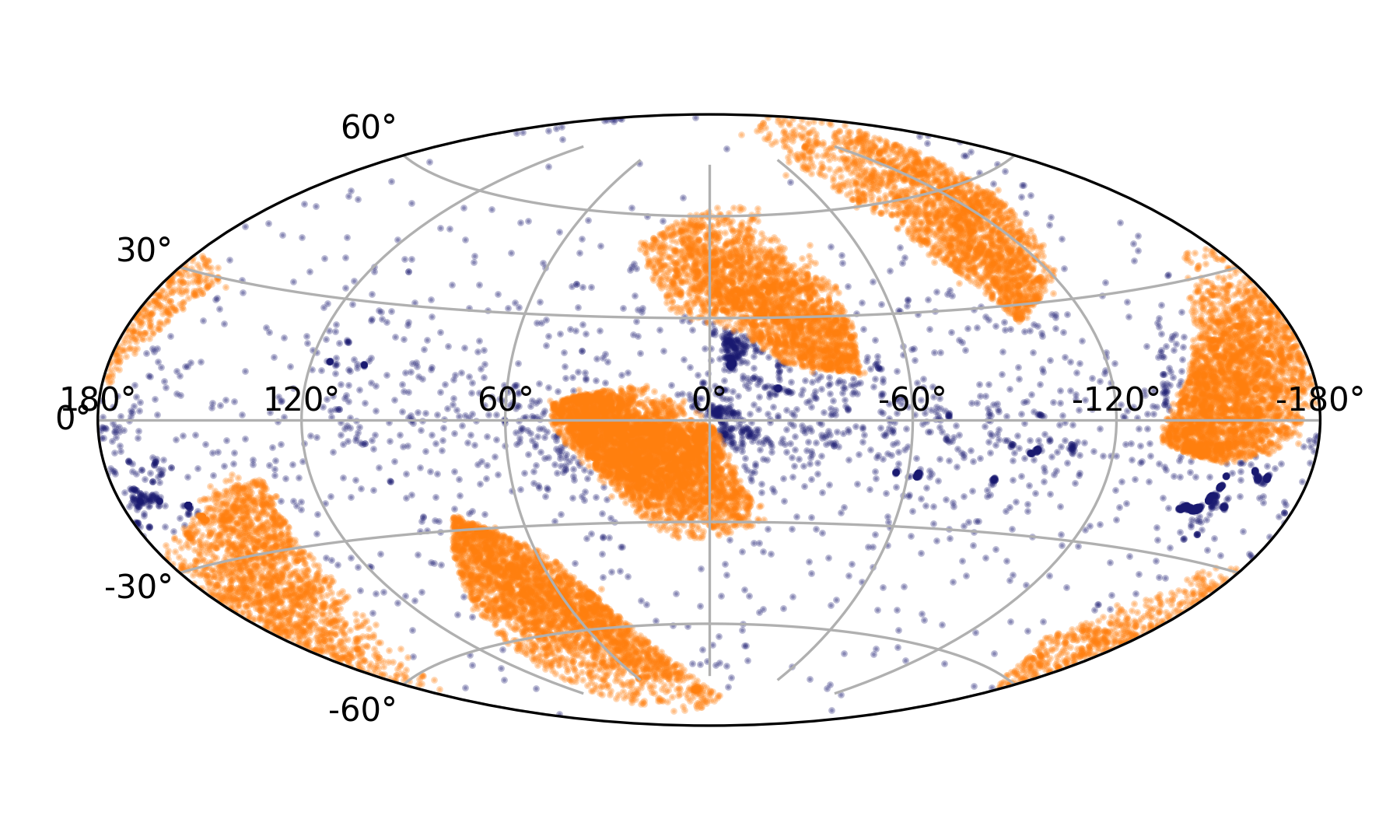}
    \caption{All sources from \textit{Gaia} DR3 within 500\,pc with pEW\,$<-0.5$\,nm but with no XP spectra publicly available. The \textit{Gaia} scanning law footprint is clearly visible. We colour-code in orange the sources that have no XP spectra because of poor \textit{Gaia} coverage. The sources in blue do not have XP spectra because of other reasons. Of the sources in orange, we only consider in criterion B (Sect. \ref{Sect:YSO_flags_criteria}) the ones with log($L_\text{acc,med-GSP-Phot extinction}/L_\odot) > -3.5$.}
    \label{fig:scanning_law}
\end{figure}

To evaluate this artificial higher-fraction of contaminants in Table \ref{table:AccretionTable_extract} in certain regions of the sky, we projected the sources with no XP spectra in the sky and manually selected the regions with no public XP spectra because of the \textit{Gaia} scanning law. Sources in the selected regions are flagged in Table \ref{table:AccretionTable_extract} in the column `flag\_Gaia\_scanning'. In selection criterion B (Sect. \ref{Sect:YSO_flags_criteria}), we keep all sources that are not in these regions, and from the sources that are in these regions we only keep the strongest accretors, defined as the sources with log($L_\text{acc,med-GSP-Phot extinction}/L_\odot) > -3.5$. This number was chosen so the density of YSO H\textalpha{} emitters in the regions with poor \textit{Gaia} coverage matches the density of YSO H\textalpha{} emitters in nearby regions with good \textit{Gaia} coverage. Fig. \ref{fig:scanning_law} shows the inhomogeneous distribution of sources with no XP spectra and the regions that we selected as affected by the \textit{Gaia} scanning law. More information on how to evaluate and work with the \textit{Gaia} scanning law can be found in \citet{2023A&A...669A..55C} and \citet{2023A&A...677A..37C}.

\section{Plots of accretion properties versus stellar parameters in Sco-Cen} \label{Appendix scocen}

\begin{figure*}[ht]
    \centering
    \begin{subfigure}[b]{\textwidth}
        \centering
        \includegraphics[width=\textwidth]{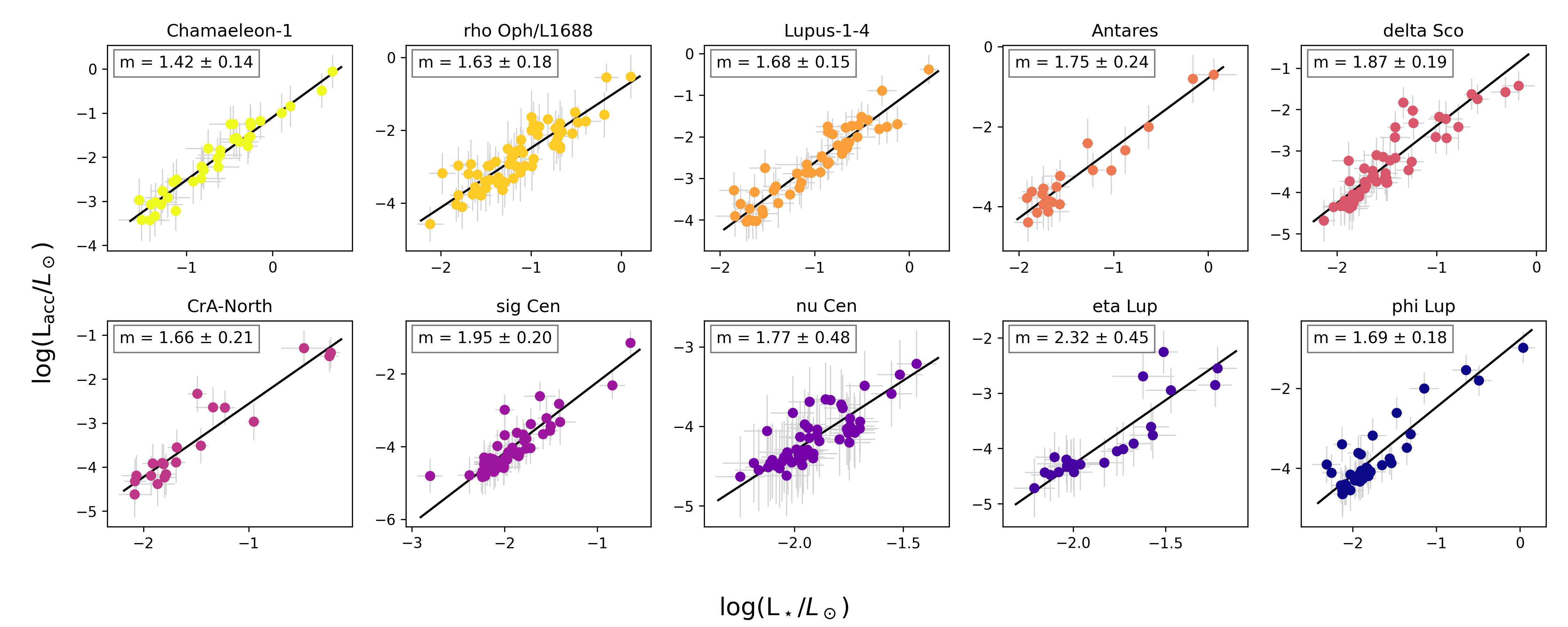} 
        \caption{}
    \end{subfigure}
    \hfill
    \begin{subfigure}[b]{\textwidth}
        \centering
        \includegraphics[width=\textwidth]{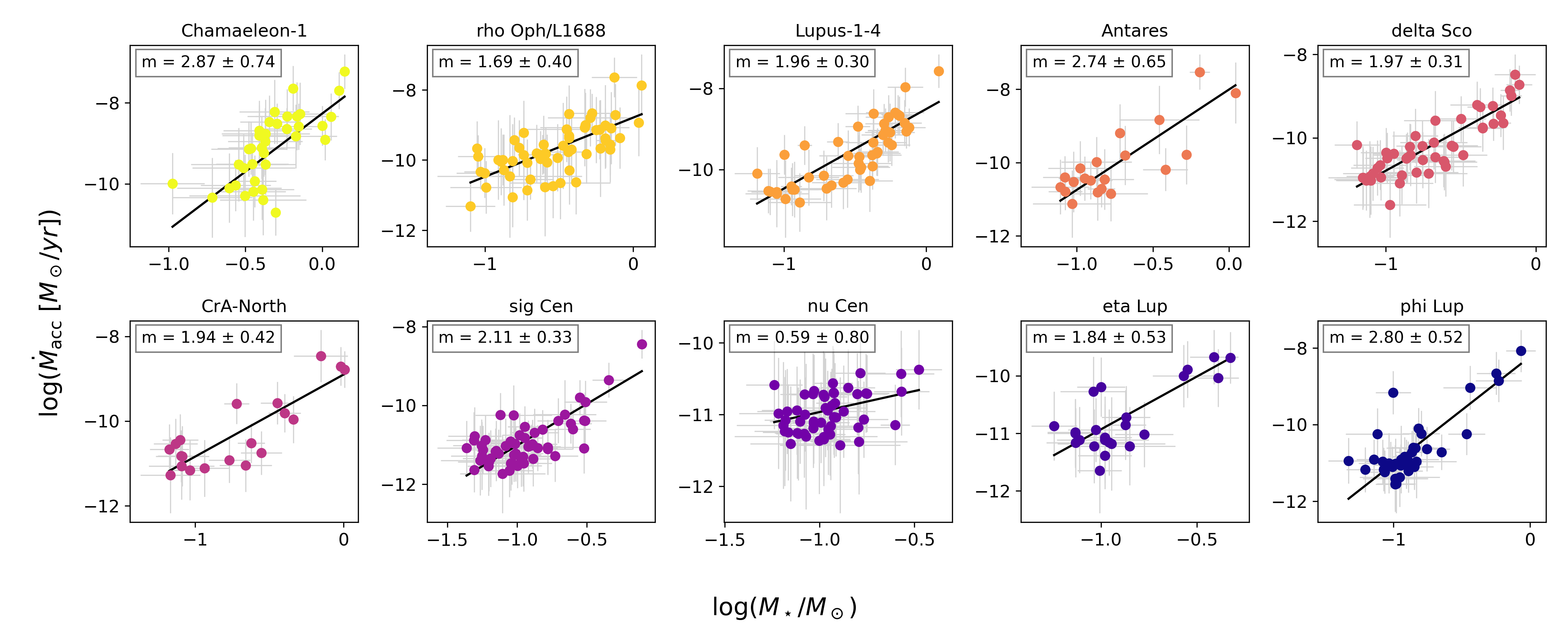}
        \caption{}
    \end{subfigure}
    
    \caption{Plots of the fits used to derive the slopes of Fig. \ref{figure_slope_change}. Each is the fit is of a specific cluster of Sco-Cen, and the colours match those used in Fig. \ref{figure_slope_change}. The value of the slope of each region is reported. Fig. (a): log $L_{\text{acc}}$ vs log $L_{\star}$; Fig. (b): log $\dot{M}_{\text{acc}}$ vs log $M_{\star}$.}
    \label{fig:sco_cen_fits}
\end{figure*}

Fig \ref{fig:sco_cen_fits} shows the fits used to derive the slopes of Fig. \ref{figure_slope_change} for ten different Sco-Cen clusters. All fits were performed using using the Bayesian method of \citet{linmix} to propagate errors.

\end{appendix}

\end{document}